\documentclass[12pt,english,longbibliography,nofootinbib,superscriptaddress,12pt,sort&compress,showkeys]{extarticle}
\usepackage[T1]{fontenc}
\usepackage[latin9]{inputenc}
\usepackage{xcolor}
\usepackage{float}
\usepackage{amsmath}
\usepackage{amssymb}
\usepackage{graphicx}
\usepackage{esint}
\usepackage[numbers]{natbib}
\PassOptionsToPackage{normalem}{ulem}
\usepackage{ulem}

\makeatletter

\providecommand{\tabularnewline}{\\}
\providecolor{lyxadded}{rgb}{0,0,1}
\providecolor{lyxdeleted}{rgb}{1,0,0}

\DeclareRobustCommand{\lyxsout}[1]{\ifx\\#1\else\sout{#1}\fi}

\usepackage{indentfirst}
\usepackage{amsfonts}
\usepackage[T1]{fontenc}
\usepackage{ae,aecompl}
\usepackage{sidecap}
\usepackage[section]{placeins}
\usepackage{epsf}
\makeatother

\usepackage{babel}
\date{\today}

\usepackage{chngcntr}

\makeatother

\usepackage{babel}
\begin{document}
\title{\textbf{Machine-learning interatomic potentials for materials science}}
\author{Y. Mishin\\
{\normalsize{}Department of Physics and Astronomy, MSN 3F3},\\
 {\normalsize{}George Mason University, Fairfax, Virginia 22030, USA}}
\maketitle
\begin{abstract}
Large-scale atomistic computer simulations of materials rely on interatomic
potentials providing computationally efficient predictions of energy
and Newtonian forces. Traditional potentials have served in this capacity
for over three decades. Recently, a new class of potentials has emerged,
which is based on a radically different philosophy. The new potentials
are constructed using machine-learning (ML) methods and a massive
reference database generated by quantum-mechanical calculations. While
the traditional potentials are derived from physical insights into
the nature of chemical bonding, the ML potentials utilize a high-dimensional
mathematical regression to interpolate between the reference energies.
We review the current status of the interatomic potential field, comparing
the strengths and weaknesses of the traditional and ML potentials.
A third class of potentials is introduced, in which an ML model is
coupled with a physics-based potential to improve the transferability
to unknown atomic environments. The discussion is focused on potentials
intended for materials science applications. Possible future directions
in this field are outlined.
\end{abstract}
\date{}


\section{Introduction}

Atomic-scale computer simulations of materials constitute a critical
component of the multiscale materials modeling paradigm \citep{HMM,Hafner00,Giessen:2020aa}.
They equip researchers with an effective tool for gaining fundamental
insights into microscopic mechanisms of processes occurring in materials
while also providing quantitative input to mesoscale and continuum
models. Modern molecular dynamics (MD)\footnote{See Appendix A for a list of abbreviations used in this article.}
and Monte Carlo (MC) simulations span length scales from a single
atom to $\sim10^{2}$ nm and time scales up to $\sim10^{2}$ ns. Access
to these length and time scales is enabled by classical interatomic
potentials (also known as classical force fields), whose role is to
predict the energy and classical forces acting on the atoms for any
given atomic configuration. Computations with classical potentials
are fast and scale linearly with the number of atoms, making them
the critical ingredient for all large-scale atomistic simulations.
The accuracy and reliability of atomistic simulations often depends
on the quality of the interatomic potentials.

The history of quantitative interatomic potentials can probably be
counted from the 1980s when the first many-body potentials for metallic
systems \citep{Daw83,Daw84} and bond-order-type potentials for covalent
materials \citep{Tersoff88,Tersoff:1988dn,Tersoff:1989wj,Stillinger85}
were introduced, and their predictive capabilities were demonstrated
in several successful applications. Since then, potentials have been
developed for most of the chemical elements (for some, several versions
are available), for many binary systems, and several ternary and higher-order
systems. Many new functional forms of the potentials have been proposed
to improve the accuracy of representing the chemical bonding in various
classes of materials.

\begin{table}[H]
\caption{Comparison of three classes of interatomic potentials.\label{tab:Comparison}}

\begin{tabular}{|l|l|c|c|}
\hline 
 & \multicolumn{3}{c|}{Potential type}\tabularnewline
\hline 
 & Traditional & ML & Physically-informed ML\tabularnewline
\hline 
\hline 
Physical foundation & Strong & None & Strong\tabularnewline
\hline 
Number of fitting parameters & $\sim10$ & $\gtrsim10^{3}$ & $\gtrsim10^{3}$\tabularnewline
\hline 
Computational speed & Very high & Slower$^{a}$ & Slower$^{a}$\tabularnewline
\hline 
Reference database & Small & Large & Large\tabularnewline
\hline 
Accuracy (interpolation) & Limited & $\sim1$ meV/atom & $\sim1$ meV/atom\tabularnewline
\hline 
Transferability (extrapolation) & Reasonable & Poor & Reasonable\tabularnewline
\hline 
Reliance on human expertise & Strong & Weaker$^{b}$ & Weaker$^{b}$\tabularnewline
\hline 
Extension to chemistries & Challenge & Challenge & Challenge\tabularnewline
\hline 
Specific to class of materials? & Yes & No & No\tabularnewline
\hline 
Systematically improvable? & No & Yes & Yes\tabularnewline
\hline 
Can be made artificial? & Yes & Maybe$^{c}$ & Maybe$^{c}$\tabularnewline
\hline 
\end{tabular}

\medskip{}

$^{a}$ but orders of magnitude faster than straight DFT calculations.

$^{b}$ Some steps of database selection and training can be partially
automatized.

$^{c}$ Not impossible in principle but we are not aware of attempts.
\end{table}

The quality of the currently available potentials varies widely. Some
reach the highest accuracy achievable with the limited number of fitting
parameters, while there is a multitude of poor-quality potentials.
Some potentials have been employed in hundreds of simulation studies
by many groups worldwide, while others were never used outside the
original publication. The construction of high-quality potentials
heavily relies on human expertise and is considered a borderline between
art and science \citep{Brenner00,Mishin.HMM}. Infrastructure has
been created for the development, testing, standardization, and storage
of interatomic potentials, including the NIST Interatomic Potentials
Repository \citep{Becker:2013aa,Hale_2018,NIST-Potentials}, the Knowledgebase
of Interatomic Models (OpenKim) \citep{Tadmor:2013aa,OpenKim}, and
potential development tools such as \emph{potfit} \citep{Brommer07,Potfit-2015,Potfit},
\emph{KLIFF} \citep{KLIFF} and \emph{Atomicrex} \citep{Stukowski:2017ui}.

Over the past decade, a new direction has emerged in this field, wherein
the interatomic potentials are constructed by machine-learning (ML)
methods, see for example \citep{Behler:2016aa,Botu:2017aa,Deringer:2019aa,Zuo:2020aa,Morawietz:2020aa,Mueller:2020vy}
for recent reviews. The idea was initially conceived in the chemistry
community\footnote{See, however, Skinner and Broughton \citep{Skinner_1995} for an early
materials science application of neural networks to construct interatomic
potentials.} in the early 1990s in the effort to improve the accuracy of intermolecular
force fields \citep{Blank:1995aa,Raff:2012aa}. After two quiet decades,
the construction of ML potentials exploded into a powerful new research
direction that gained popularity in computational materials science,
computational physics, and computational chemistry. In materials science,
the emergence and development of the ML potentials can be seen as
part of the more general quest for data-driven approaches capable
of accelerating the discovery and design of new materials \citep{Mueller:2016aa,Ramprasad:2017aa,Gubernatis:2018aa,Rickman:2019aa,Ong:2019aa,Pablo:2019aa,Picklum:2019aa,DeCost_2020,Saal:2020aa,Batra:2020aa}.
In simple terms, the basic idea of ML potentials is to forego the
physical insights and try to predict the potential energy of the system
by numerical interpolation between known reference data (energy, forces
and often stresses) generated by quantum-mechanical calculations.
This approach signifies a radical departure from the traditional potentials
aiming to achieve the same goal by capturing the basic physics of
interatomic bonding in the material in question.

The goal of this article is to review the current status and offer
the author's view of the future of the interatomic potential field.
Attention will be focused on potentials intended for materials science
applications, such as the modeling of microstructure, defects, mechanical
and thermal properties, and alloy thermodynamics and kinetics. This
excludes the numerous chemical applications, molecular matter, and
molecule-surface interaction systems. The reader interested in ML
force fields for molecular systems is referred to the recent literature
\citep{Bereau:2015,Behler:2016aa,Schutt:148aa,Bereau:2018aa,Cooper:2020aa}.
In terms of the ML methods, we mostly discuss the high-dimensional
regression models, which are utilized for the potential training and
fall in the category of supervised ML. Classification problems, pattern
recognition, clustering, and many other unsupervised learning approaches
\citep{Mueller:2016aa,Gubernatis:2018aa,Rickman:2019aa,Ong:2019aa,Pablo:2019aa,DeCost_2020}
are also actively used in materials research but lie outside the scope
of the article.

The leading theme of the article is the comparison of the traditional
and ML potentials. We discuss their strong and weak points, some of
which are complementary to each other. A summary of this comparison
is presented in Table \ref{tab:Comparison}, with a more detailed
analysis to follow. After a brief overview of the traditional potentials
in section \ref{sec:Traditional-potentials}, we review the general
idea of the ML potentials (section \ref{subsec:Basic-idea-ML}) followed
by a more detailed discussion of the technical aspects, such as the
types of regression, the structural descriptors, the reference database
construction, and the training process. A summary of the ML potentials
is presented in section \ref{subsec:Discussion-of-ML}. Section \ref{sec:Physically-informed-ML}
introduces the general idea of the physically-informed ML potentials
combining the high training accuracy with physics-based transferability.
The approach is illustrated by the specific example of physically-informed
neural network potentials. Finally, in section \ref{sec:Summary}
we summarize this review and present our view of the history and vision
of the future of this field.

\section{The traditional interatomic potentials\label{sec:Traditional-potentials}}

\subsection{What are the potentials, and why do we need them?}

The most accurate energy and force calculations are performed by electronic
structure methods based on the direct quantum-mechanical treatment
of the electrons. Since the density functional theory (DFT) \citep{hohenberg64:dft,kohn65:inhom_elec}
is employed in most of such calculations, we will refer to them for
brevity as ``DFT calculations''. In addition to the high accuracy
(typically, a few meV/atom) and deep physical underpinnings, the DFT
calculations can be applied to both elemental and multicomponent systems
with nearly equal computational effort. This makes DFT calculations
a highly effective tool for a broad exploration of materials chemistry
\citep{Curtarolo:2012qf,Curtarolo:2013aa,Walle:2019aa}. DFT calculations
also provide access to a broad spectrum of physical properties, ranging
from mechanical to electronic, magnetic and optical. A major limitation
of the DFT calculations is that they are computationally demanding
and scale with the number of atoms $N$ as $N^{3}$ or slower. At
present, static DFT calculations can only be performed for systems
containing a few hundred atoms. \emph{Ab initio} molecular dynamics
(AIMD) can be run for about a hundred picoseconds.

Meanwhile, the intrinsic length and time scales of many materials
processes greatly exceed the scales currently accessible by DFT calculations.
Examples include plastic deformation, fracture, phase nucleation and
growth (including, for example, alloy melting and solidification),
and the microstructure evolution by solid-solid interface migration.
The modeling of such processes requires access to large collections
of atoms and statistical averaging over multiple thermally activated
events. Classical interatomic potentials offer a solution by enabling
drastically accelerated MD and MC simulations at the price of significantly
reduced accuracy. In addition to the accuracy compromise, the classical-mechanical
nature of the potential-based simulations excludes any treatment of
electric, magnetic or optical properties.

\begin{figure}[h]
\noindent \begin{centering}
\includegraphics[width=0.5\textwidth]{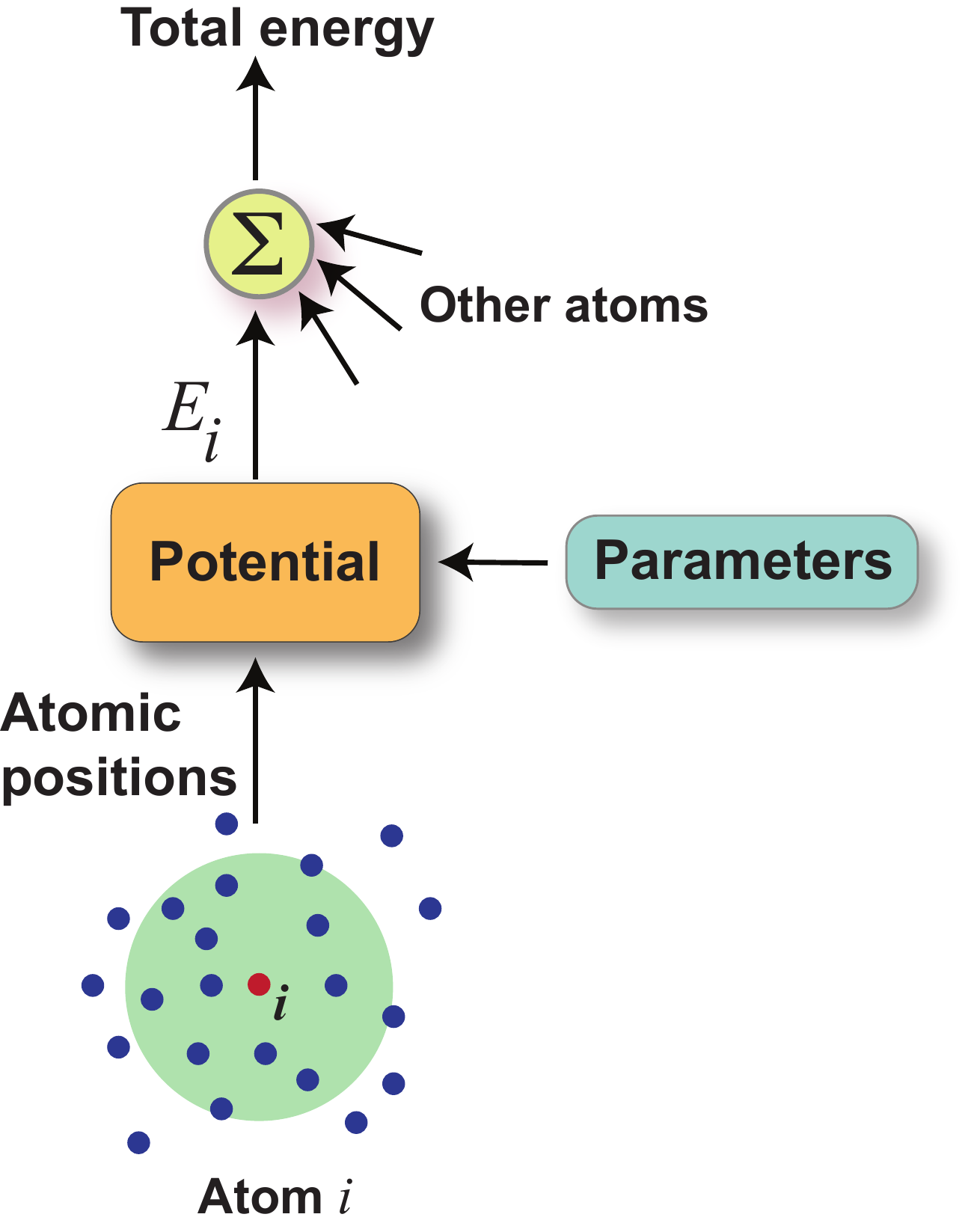}
\par\end{centering}
\bigskip{}

\caption{Flowchart of total energy calculations with traditional interatomic
potentials. The energy $E_{i}$ of an atom $i$ is computed using
atomic coordinates within the cutoff sphere (green) and fixed values
of the potential parameters. The atomic energies of all atoms of the
system are summed up (symbol $\Sigma$) to obtain the total energy.\label{fig:Trad_pots}}
\end{figure}

Interatomic potentials parameterize the system's configuration space
and express its potential energy $E$ as a function of the atomic
positions (Fig.~\ref{fig:Trad_pots}). This function can be represented
by a $3N$-dimensional hypersurface\footnote{For simplicity, we consider an elemental system. For multicomponent
systems, the configuration space additionally includes permutations
of different chemical species.} called the potential energy surface (PES). Knowing the PES, the forces
$\mathbf{F}_{i}=-\partial E/\partial\mathbf{r}_{i}$ acting on individual
atoms $i$ can be computed for any atomic configuration ($\mathbf{r}_{i}$\ being
the position vector of atom $i$). Almost all potentials partition
the total energy into energies $E_{i}$ assigned to individual atoms:
$E=\sum_{i}E_{i}$. Each atomic energy $E_{i}$ is expressed as a
function of the atomic positions $\mathbf{R}_{i}\equiv(\mathbf{r}_{i1},\mathbf{r}_{i2},...,\mathbf{r}_{in_{i}})$
in the vicinity of the atom. The functional form of the potential,
\begin{equation}
E_{i}=\Phi(\mathbf{R}_{i},\mathbf{p}_{i}),\label{eq:1}
\end{equation}
ensures the invariance of the energy under rotations and translations
of the coordinate axes and permutations of the atoms. In Eq.(\ref{eq:1}),
$\mathbf{p}_{i}$ is a set of parameters discussed below. The partitioning
into atomic energies accelerates the total energy calculation by making
it a linear $N$ procedure and enabling parallelization by domain
decomposition. We emphasize, however, that this partitioning is only
valid for systems with short-range interactions. Long-range Coulomb
and dispersive interactions must be added as separate terms and computed
by the Ewald summation or similar numerical methods.

\subsection{The physical basis of traditional potentials}

The distinguishing feature of the traditional potentials is that the
potential function $\Phi(\mathbf{R}_{i},\mathbf{p}_{i})$ is based
on a physical understanding of interatomic bonding in the material
(Table \ref{tab:Comparison}). For example, the embedded-atom method
(EAM) \citep{Daw83,Daw84,Finnis84}, the modified EAM (MEAM) \citep{Baskes87},
and the angular-dependent potential (ADP) \citep{Mishin05a} are specifically
designed for metallic systems. The Tersoff \citep{Tersoff88,Tersoff:1988dn,Tersoff:1989wj}
and Stillinger-Weber \citep{Stillinger85} potentials were specifically
developed for strongly covalent materials such as silicon and carbon.
The charge-optimized many-body (COMB) potentials \citep{Liang:2012aa},
reactive bond-order (REBO) potentials \citep{Brenner90,Brenner00,Stuart:2000aa},
and reactive force fields (ReaxFF) \citep{van-Duin:2001aa} are most
appropriate for molecular systems with chemical reactions. The functional
forms of the traditional potentials are diverse and largely incompatible
with each other due to the differences in the underlying physical
and chemical models specific to the respective classes of materials.
This disparity of the functional forms poses a formidable challenge
to modeling mixed-bonding and two-phase systems containing metal-ceramic
or metal-polymer interfaces.\footnote{Technically, one can always create \emph{ad hoc} functional forms
of cross-element interactions that mathematically reduce to the respective
single-element functions for certain combinations of parameters, as
recently proposed for metal-semiconductor systems \citep{Dongare:2009aa,Dongare:2009ab,Saidi_2014}.
Such functions are not motivated by physical insights, and their reliability
is likely to be limited.}

\subsection{Training of traditional potentials}

The potential function (\ref{eq:1}) of the traditional potentials
depends on a small number $m$ (usually, 10 to 20) of global (same
for all atoms) fitting parameters $\mathbf{p}=(p_{1},...,p_{m})$.
These parameters are optimized by training on a database usually composed
of experimental data and a relatively small number of DFT energies
or forces. The experimental information comes in the form of specific
physical properties of the targeted material. Such properties typically
include the lattice constant, cohesive energy, elastic constants,
point defect formation and migration energies, surface energies, and
generalized stacking fault energies. By contrast to the ML potentials
discussed later, the traditional potentials are fitted \emph{directly}
to these properties, not the PES. Once optimized, the potential parameters
are fixed once and for all and are used for predicting the energy
and forces in all atomic configurations encountered during the subsequent
simulations. Due to the mathematical simplicity of the potential function,
such calculations are computationally fast, easily parallelizable,
and provide access to systems containing millions of atoms.

The loss function minimized during the potential training is usually
the mean squared deviation of properties from their reference values.
These deviations are included with weights assigned to individual
properties and playing the role of hyper-parameters. Since the reference
database is small, a single training run is computationally fast.
However, the potential obtained has to be tested for many physical
properties not represented in the training database. Some of the tests,
such as computing the melting temperature, require lengthy simulations.
The optimization process has a feedback loop in which the hyper-parameters
are adjusted by the developer to improve the testing results. This
loop is the most critical step of the potentials construction that
heavily relies on human decisions and can hardly be automatized. Relationships
between the weights of the fitted properties and the accuracy of the
tested properties are not apparent. Decisions have to be made based
on the developer's prior experience, intuition, and knowledge of many
intricacies of atomistic simulations. The enormous complexity of the
property-based optimization and the reliance on expert knowledge make
the development of high-quality potentials a long and painful process.

The accepted practice in constructing multicomponent potentials is
to preserve the underlying elemental potentials and only fit the parameters
of the cross-interaction functions. With this strategy, one elemental
potential can be crossed with many others. This property of multicomponent
potentials, which we call the ``inheritance'' of the elemental potentials,
helps avoid duplication of potentials and greatly facilitates their
standardization and organization in repositories. Nevertheless, the
chemical exploration using potentials is much more complicated than
with DFT calculations. Each time an element must be added to the system,
a new set of cross-interaction functions must be fitted, which requires
significant efforts.

\subsection{Accuracy and transferability of traditional potentials\label{subsec:trand-poten-4}}

Although the construction of traditional potentials is based on physical
insights, the underlying physical models are highly approximate and
contain few adjustable parameters. As a result, their accuracy is
rather limited (Table \ref{tab:Comparison}). While some properties
are reproduced with decent accuracy, there are many subtle effects
(such as specific surface reconstructions or complex dislocation core
structures) that are not predicted correctly \citep{Lysogorskiy:2019aa}.
Traditional potentials are especially struggling with complex elements
such as Si capable of exhibiting both covalent and metallic types
of bonding. Carbon is another difficult case due to its bonding complexity
and the existence of multiple metastable 3D, 2D, and 1D structures.
There are many examples of wrong predictions by traditional potentials
and their failures to describe particular properties in particular
systems. Nevertheless, many predictions were subsequently confirmed
by experiment or DFT calculations. Much of the current knowledge of
dislocations, grain boundaries, and other interfaces emerged from
potential-based MD and MC simulations performed over the past decades.

\begin{figure}
\noindent \begin{centering}
\includegraphics[width=0.35\textwidth]{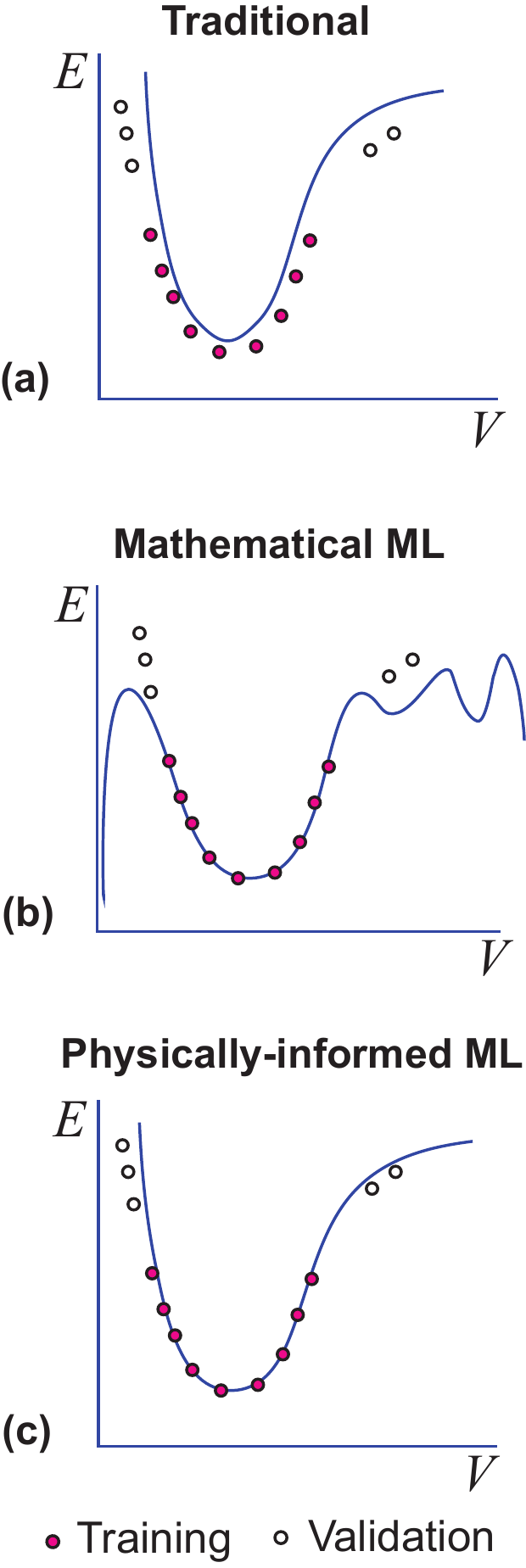}
\par\end{centering}
\caption{Schematic illustration of accuracy and transferability of (a) traditional
(b) mathematical ML and (c) physically-informed ML interatomic potentials.
The energy-volume ($E$-$V$) relation for a particular structure
obtained by DFT calculations (points) is compared with predictions
of the potentials. The points inside and outside the training domain
are shown by filled and open circles, respectively.\label{fig:Comparison of potentials}}
\end{figure}

Despite the limited accuracy, the traditional potentials often demonstrate
reasonably good transferability to atomic configurations lying well
outside the training dataset.\footnote{Transferability of potentials is often interpreted as their ability
to make accurate predictions for structures not included in the reference
dataset. Here, we understand this term as the potential's ability
to make meaningful predictions \emph{outside} the domain of reference
structures, i.e., in the \emph{extrapolation} regime. There are several
different criteria for distinguishing extrapolation from interpolation
(based on either statistical uncertainties or definitions of distances
and convexity in the feature space). However, in many cases the distinction
is quite obvious. For example, testing a potential for densities lying
well outside the density range represented in the reference dataset
can be considered extrapolation.} This important property of traditional potentials is due to the incorporation
of basic physics in their functional form (Table \ref{tab:Comparison}).
As long as the nature of the chemical bonding in the material remains
the same as was assumed during the potential construction, the potential
should make at least physically meaningful energy predictions for
new configurations not represented during the training (Fig.~\ref{fig:Comparison of potentials}a).

\subsection{Classification of interatomic potentials}

In terms of intended applications, all potentials, regardless of their
functional form, can be classified into three categories: general-purpose
type, special-purpose type, and artificial.

\emph{The general-purpose potentials} are trained to reproduce a broad
spectrum of physical properties considered most important for the
subsequent atomistic simulations. Although the training process targets
a particular set of properties, the underlying reference structures
must be diverse enough to represent the most typical atomic environments
occurring in typical simulations. Once released to the community,
a general-purpose potential is used for almost any type of simulation
that the user may choose to perform. Most of the atomistic simulations
conducted today utilize such off-the-shelf potentials. In addition
to the computational speed, the reason for their popularity is their
reasonable transferability mentioned above. Even when taken well outside
the training domain, the potential may lose much of its accuracy but
does not usually generate physically nonsensical results.

\emph{Special-purpose potentials} are designed for one particular
type of simulation. They target a specific application and are not
expected to be transferable to other types of simulation. For example,
a potential can be specifically trained to reproduce the lattice dynamics
and phonon thermal conductivity of a particular element or compound.

The third class is comprised of the so-called \emph{artificial (synthetic)
potentials}. Taking advantage of the property-based training, one
can construct a series of artificial potentials with a varying value
of a particular physical property (or properties) while keeping other
properties unaltered. Simulations with such potentials and comparison
with experiments or theory may help the user better understand the
impacts of the different physical parameters on a particular process.
For example, one can generate a set of face-centered-cubic (FCC) potentials
with a varying stacking fault or twin boundary energy with all other
properties fixed. Simulation of plastic deformation with these potentials
may help disentangle the effects of the fault energies on the deformation
modes from other possible effects. As another example, simulations
with artificial Al potentials that significantly modified the liquid
properties helped the authors \citep{Mendelev_2010} to unravel a
relationship between the solid-liquid interface mobility and the liquid
diffusion coefficient. Artificial potentials are also part of the
simulated alchemy approach \citep{Broughton83b,Straatsma:1992aa,Skinner95},
in which a reversible path between two states is implemented by tampering
with the potential. For example, the potential can be modified by
small increments (e.g., by the rule of mixtures) to transform one
elemental potential into another. Alternatively, some atoms can be
slowly removed from the system (made invisible) by gradually dialing
down their interactions with the remaining atoms. The free energy
difference between the two states of the system can be then computed
by the $\lambda$-integration method \citep{Broughton83b,Lill:1997aa,Frenkel02,Addula:2020aa}.

We emphasize that the above classification is based on the intended
usage of the potential and is common to both traditional and ML potentials.
The same functional form can be used to generate a potential for any
of the three categories.

\section{Machine-learning potentials\label{sec:ML}}

\subsection{The basic idea\label{subsec:Basic-idea-ML}}

While the traditional potentials target a particular set of properties,
the ML potentials map the $3N$-dimensional configurational space
of the system onto its PES. The latter is represented by a discrete
set of DFT energies included in the training dataset. The mapping
is implemented by a numerical interpolation algorithm (regression)
containing a large number of adjustable parameters. The goal of the
training is to optimize the regression parameters to obtain a smooth
PES that best interpolates between the reference energies. The energy
gradients at the reference points, obtained from the DFT atomic forces,
can also be included in the optimization process. Given the large
size of the reference database and the high dimensionality of the
parameter space, the optimization problem is complex and greatly benefits
from the application of ML methods.

Like the traditional potentials, most of the ML potentials are predicated
on the locality of atomic interactions and thus partition the total
energy $E=\sum_{i}E_{i}$ into atomic energies $E_{i}$. Considering
an elemental material to simplify the discussion, the local environment
of an atom $i$ is defined by the set of positions $\mathbf{R}_{i}\equiv(\mathbf{r}_{i1},\mathbf{r}_{i2},...,\mathbf{r}_{in})$
of its $n$ neighbors within a cutoff sphere of a radius $r_{c}$
(for a multicomponent system, the chemical identities of the atoms
are also considered). The local position vector $\mathbf{R}_{i}$
is mapped onto the local energy by the potential function (\ref{eq:1}).
The total PES is reconstructed by the summation of such local maps.
As with the traditional potentials, the locality approximation accelerates
the total energy calculation and enables its effective parallelization
by the spatial domain decomposition of the system. The locality also
justifies using DFT calculations for small supercells to make the
energy predictions for large systems. (Efforts to include long-range
interactions due, for example, to electrostatic forces have also been
published \citep{Artrith:2011aa,Artrith:2011ab,Unke:2019wf}.)

\begin{figure}
\noindent \begin{centering}
\includegraphics[width=0.4\textwidth]{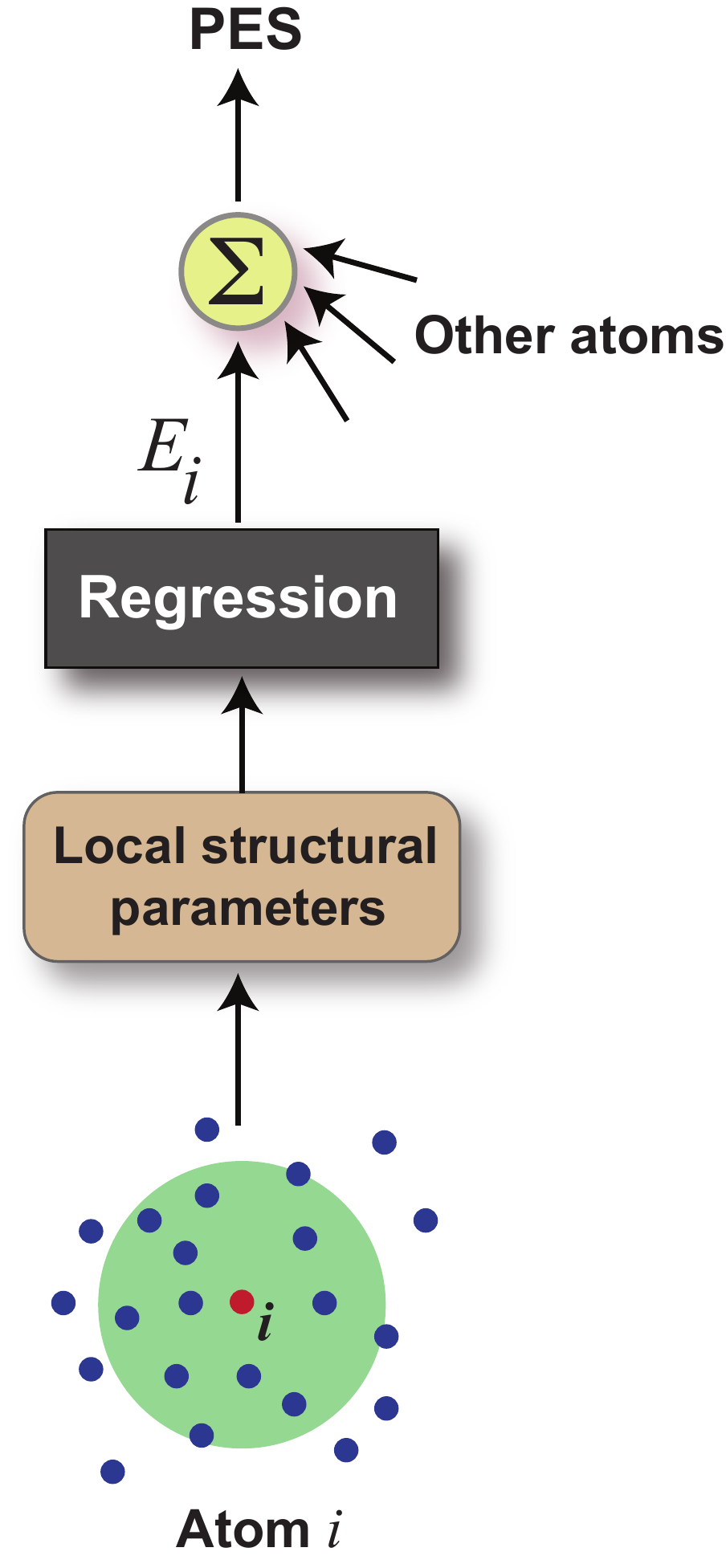}
\par\end{centering}
\bigskip{}

\caption{Flowchart of total energy calculations with ML interatomic potentials.
The local environment of an atom $i$ within the cutoff sphere (green)
is encoded in a set of local structural parameters, which are then
mapped onto the energy $E_{i}$ assigned to atom $i$ using a regression
model. The summation of the energies of other atoms of the system
(symbol $\Sigma$) gives the total energy and thus a point on the
PES of the system. \label{fig:ML_pots-1}}
\end{figure}

The local mapping is implemented in two steps. First, instead of the
position vector $\mathbf{R}_{i}$, the local atomic environment is
represented by another vector composed of local structural parameters
$\mathbf{G}_{i}=(G_{i1},G_{i2},...,G_{iK})$. These parameters are
smooth functions of $\mathbf{R}_{i}$ invariant under translations
and rotations of the coordinate axes and permutations (relabeling)
of the atoms. At the second step, the vector $\mathbf{G}_{i}$ is
mapped onto the energy $E_{i}$ by a chosen regression model $\mathcal{R}$.
Thus, the atomic energy calculation can be represented by the formula
\begin{equation}
\mathbf{R}_{i}\rightarrow\mathbf{G}_{i}\overset{\mathcal{R}}{\rightarrow}E_{i}\label{eq:ML-1}
\end{equation}
shown diagrammatically in Fig.~\ref{fig:ML_pots-1}.

The role of the structural descriptors $\mathbf{G}_{i}$ is twofold.
First, they ensure the mentioned invariance and smoothness of the
PES. The second role of the $\mathbf{G}_{i}$'s is to replace the
variable-size position vector $\mathbf{R}_{i}$ (whose length $n$
can vary from one atom to another according to the number of neighbors)
by a feature vector of a \emph{fixed} length $K$. The introduction
of a fixed number of local structural descriptors was a crucial step
proposed by Behler and Parrinello \citep{Behler07}. Although they
initially focused on a single-component system, the general idea of
fixing the size of the descriptors was later extended to multicomponent
systems \citep{Artrith:2011aa,Artrith:2011ab,Sosso2012,Artrith:2016aa,Artrith:2017aa,Hajinazar:2017aa,Kobayashi:2017aa,Artrith:2018aa,Li:2018aa,Hajinazar:2019aa,Gubaev:2019aa,Zhang:2019ab,Andolina:2020aa}.
With $K$ fixed, the total energy calculation can be accomplished
with a \emph{single} pre-trained regression $\mathcal{R}$ mapping
the $K$-dimensional feature space onto the 1D space of atomic energies.
(This explains why the regression symbol $\mathcal{R}$ in Eq.(\ref{eq:ML-1})
does not carry the index $i$.) The possible regression models $\mathcal{R}$
implementing this mapping will be discussed later.

The preceding discussion points to three distinguishing features of
ML potentials compared with the traditional potentials (Table \ref{tab:Comparison}):
\begin{itemize}
\item The reference database is generated by DFT calculations without any
experimental input.
\item The energy is predicted by purely numerical interpolation of the reference
dataset without any physics-based model. The only physical input is
the assumption of the locality of the atomic interactions and the
invariance of energy under translations, rotations, and permutations
of atoms.
\item The potential is trained to approximate the PES of the system and
not a particular physical property or properties.
\end{itemize}
In principle, the PES uniquely defines all properties of the system.
Its local minima represent stable or metastable structures; the saddle
points correspond to energy barriers of thermally activated kinetic
processes, such as vacancy jumps or Peierls barriers of dislocations,
while the PES curvature controls the elastic constants and phonon
dispersion relations. In reality, however, the inevitable deviations
of the approximate PES from the theoretical one obtained by DFT calculations
cause errors in the predicted physical properties. We will return
to the PES versus properties issue later. In the following sections,
we briefly review some of the specific steps of the ML potential development.

\subsection{The local structural descriptors\label{subsec:Descriptors-ML}}

The local structural parameters $\mathbf{G}_{i}=(G_{i1},G_{i2},...,G_{iK})$
encode the local environment of every atom $i$ in a fixed number
of invariant parameters, often referred to as local fingerprints.
As already mentioned, the underlying assumption is that the atomic
interactions are short-range, and thus the energy assigned to atom
$i$ only depends on its local environment. The total energy is predicted
based on the information about all local environments in the system.

The vector $\mathbf{G}_{i}$ is a function of the neighbor positions
$\mathbf{R}_{i}$ (and their chemical identities in multicomponent
systems). The choice of this function is extremely important and can
strongly impact the accuracy of the potential. The general requirement
for this function is to capture the local atomic environment most
efficiently. The efficiency includes the resolution (different environments
must be represented by sufficiently different descriptors), the descriptor
size $K$, and the computational cost of its calculation. A more detailed
list of expected properties of descriptors can be found in \citep{Onat:2020wk}.
It is also desirable that the set of descriptors be complete, i.e.,
capable of exactly reconstructing the local environment (up to symmetry
operations) at least in principle. In the absence of completeness,
the descriptors can miss certain structural features. On the other
hand, an overcomplete set can generate different descriptors for the
same structure, resulting in discontinuous behavior of energy predictions.
Several representations based on complete basis sets have been proposed
\citep{Drautz:2019aa,Kocer:2020wn}. In practice, only two- and and
three-body terms are usually utilized. The issues surrounding the
incompleteness of such ``truncated'' representations have been recently
investigated \citep{Pozdnyakov:2020tw}.

More detailed theoretical analysis and benchmarking of different descriptors
can be found in the literature \citep{Bartok:2013aa,Batra:2019aa,Willatt:2019aa,Drautz:2019aa,Zuo:2020aa,Himanen:2020aa,Kocer:2020wn,Pozdnyakov:2020tw,Onat:2020wk}.
In practice, different authors use their favorite $\mathbf{G}_{i}$'s,
which often perform equally well. Some of the commonly used descriptors
include:

\emph{Gaussian descriptors.} Combinations of two-body and three-body
Gaussian functions of interatomic distances and bond angles are multiplied
by a smooth cutoff function. A particular case, called the symmetry
functions, was proposed by Behler and Parrinello \citep{Behler07},
but other functional forms can be equally efficient \citep{Botu:2017aa,Batra:2019aa,PINN-1,Pun:2020aa,Ta-PINN-in-review}.
It was proposed \citep{PINN-1,Pun:2020aa,Ta-PINN-in-review} to express
the bond-angle dependence through Legendre polynomials since they
form an orthogonal and complete set.

\emph{Zernike descriptors.} The atomic environment is represented
by coefficients (moments) on the basis of Zernike functions \citep{Novotni:2004aa,Khorshidi:2016aa},
which have the advantage of forming an orthogonal basis set and automatically
including many-body interactions. Khorshidi et al.~\citep{Khorshidi:2016aa}
demonstrate that the Zernike descriptors are computationally faster
than the bispectrum method (see below) and that their derivatives
(needed for the force calculations) can be computed more easily.

\emph{Moment tensor descriptors} \citep{Shapeev:2016aa}. Moment tensors
of different ranks are formed by multiplying radial functions by outer
products of the position vectors of the neighboring atoms. Rotationally
and permutationally invariant descriptors are obtained from contractions
of these tensors to produce scalars. These descriptors are used as
part of the moment tensor potentials (MTP) \citep{Shapeev:2016aa,Podryabinkin:2017aa,Podryabinkin:2019aa,Novikov:2019aa,Novikov:2021aa}.
The MTP descriptors form a complete basis set of polynomials if all
orders are included. In the recent tests of several ML potentials
\citep{Zuo:2020aa}, the MTP potentials have shown the optimal combination
of accuracy and computational efficiency.

\emph{Smooth overlap of atomic positions (SOAP)} \citep{Bartok:2010aa,Bartok:2013aa}.
The neighboring atoms are represented by overlapping Gaussian peaks
of density, which are expanded in spherical harmonics. The bispectrum
descriptors are formed from the expansion coefficients and are rotationally
invariant. The SOAP descriptors have proved to be very powerful but
computationally slower than the alternatives mentioned above and below.
Although designed in tandem with the Gaussian approximation potentials
(GAP), they can also be combined with neural networks \citep{Khorshidi:2016aa}
and other regression models.

\emph{Spectral neighbor analysis potential (SNAP) descriptors} \citep{Thompson:2015aa}.
The fingerprinting of the environment is somewhat similar to that
in the SOAP method. The density peaks corresponding to the neighboring
atoms are expanded on the basis of 4D hyper-spherical harmonics. The
bispectrum formed by the expansion coefficients provides the local
structural descriptors. A multicomponent extension of the method has
been recently developed \citep{Cusentino:2020uc}.

\emph{Atomic cluster expansion (ACE)} \citep{Drautz:2019aa} can be
viewed as a generalization of some of the above descriptors. The atomic
environment is represented by a set of invariant polynomials of functions
forming a complete basis. Each basis function is a product of a radial
function and an angular component represented by a spherical harmonic.
Linear scaling with the number of neighbors is achieved by transforming
the sums products into products of sums (which is also done with the
MTP descriptors)\footnote{See \citep{Seko:2019vu} for a general analysis of rotational invariants
based on spherical harmonics.}. The expansion can be applied to multicomponent environments. For
atomic clusters, ACE provides a good linear model for energy predictions.
For bulk systems, a nonlinear model has been proposed by combining
the ACE descriptors with a traditional interatomic potential in the
Finnis-Sinclair \citep{Finnis84} or any other format. The method
has been recently extended to vectorial and tensorial physical properties
\citep{Drautz:2020tl}.

Assessing the efficiency of descriptors is challenging since their
performance depends on the regression model and the database used
for the testing. The benchmarking of different potentials \citep{Zuo:2020aa}
does not shed much light on the descriptors' performance since they
are only responsible for one step in the energy calculation. Onat
et al.~\citep{Onat:2020wk} have recently published a detailed comparison
of the sensitivity of eight most popular descriptors to structural
perturbations using four different reference datasets.

\subsection{The regression models\label{subsec:Regression-ML}}

Several high-dimensional regression models are available for mapping
the local environments of atoms onto the PES. The most common choices
are the Gaussian process regression \citep{Payne.HMM,Bartok:2010aa,Bartok:2013aa,Li:2015aa,Glielmo:2017aa,Bartok_2018,Deringer:2018aa}
(underlying the GAP potentials), the kernel ridge regression \citep{Botu:2015bb,Botu:2015aa,Mueller:2016aa},
the SNAP model \citep{Thompson:2015aa,Chen:2017ab,Li:2018aa}, the
MTP potentials \citep{Shapeev:2016aa}, and the artificial neural
network (NN) regression \citep{Behler07,Bholoa:2007aa,Behler:2008aa,Sanville08,Eshet2010,Handley:2010aa,Behler:2011aa,Behler:2011ab,Sosso2012,Behler:2015aa,Behler:2016aa,Schutt:148aa,Imbalzano:2018aa}.
Some of the models (such as SNAP and MTP) are linear, while others
are highly nonlinear. They contain a large ($\sim10^{3}$) number
of parameters, which are trained on a DFT database.

We will concentrate the discussion on the NN regression as an example.
NNs have the advantage of being very flexible, not affected by any
constraints specific to the material system, and being universal approximators
\citep{Kurkova:1992aa}. NNs are widely used in materials science
and many other areas of science and technology. Thus, the existing
experience, methodologies, and even some software packages can be
transferred to the potential development field.

\begin{figure}
\noindent \begin{centering}
\includegraphics[width=0.77\textwidth]{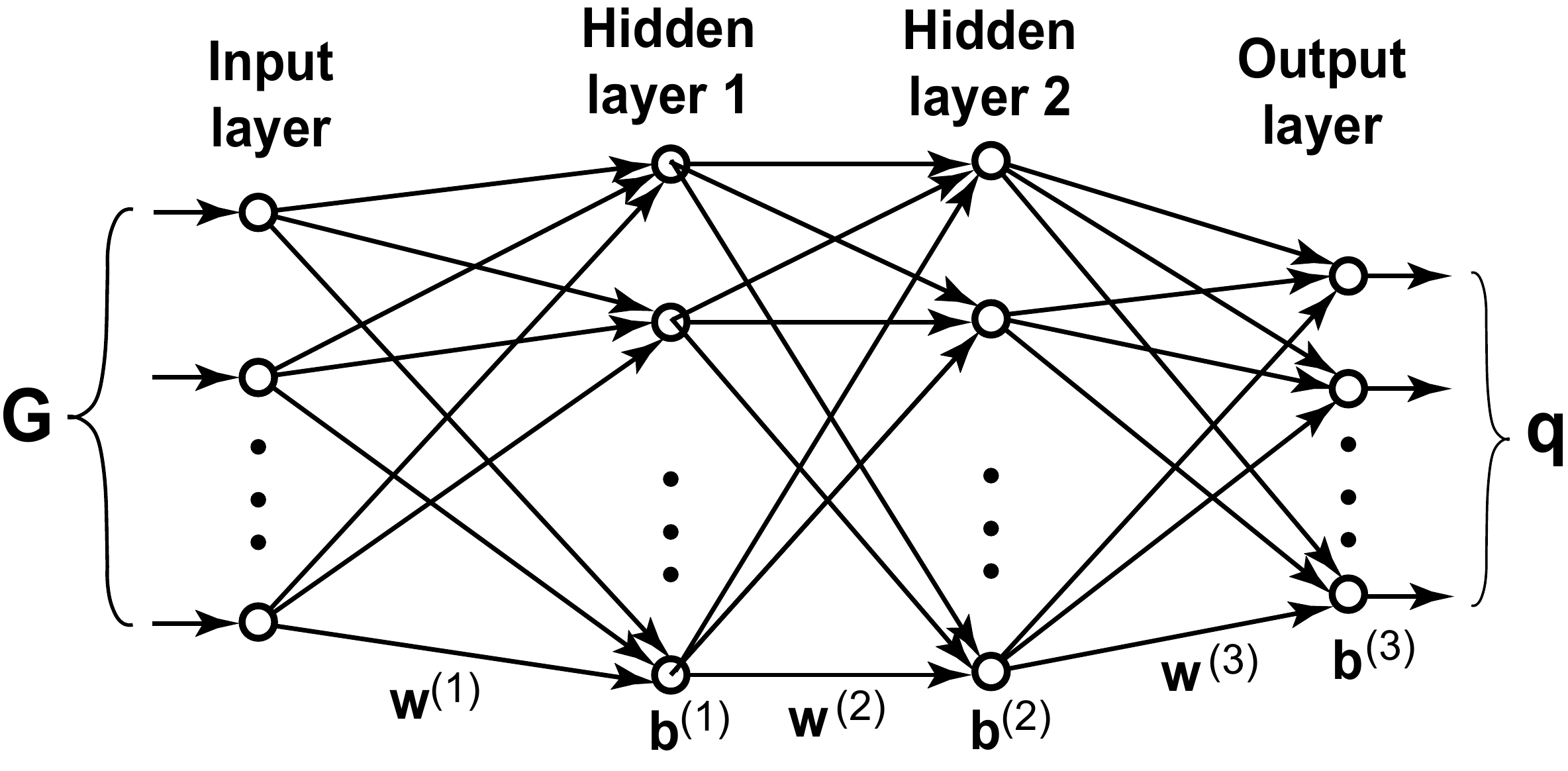}
\par\end{centering}
\caption{Example of a feed-forward NN containing two hidden layers. $\mathbf{G}$
is the input vector, $\mathbf{q}$ is the output vector. The signals
transmitted between the NN nodes (neurons) are transformed by the
weight matrices $\mathbf{w}^{(1)}$, $\mathbf{w}^{(2)}$ and $\mathbf{w}^{(3)}$
and the bias vectors $\mathbf{b}^{(1)}$, $\mathbf{b}^{(2)}$ and
$\mathbf{b}^{(3)}$. \label{fig:NN-diagram}}
\end{figure}

Most of the NN potentials utilize a simple feed-forward architecture,
where the nodes (neurons) are organized in layers. The feature vector
is fed into the input layer, the output layer delivers the energy
or force, and the hidden layers inserted in between provide additional
adjustable parameters and enhance the flexibility of the model (Fig.~\ref{fig:NN-diagram}).
The output layer consists of just one node for PES fitting, but three
more nodes with Cartesian components of the force can be added if
force matching is part of the training. Generally, we can consider
a feed-forward NN with an architecture $K-k-l...-m$ composed of $M$
layers labeled by an index $n=1,2,...,M$. The input layer ($n=1$)
contains $K$ nodes, which receive the input parameters $G_{\eta}$,
$\eta=1,2,..,K$. These are multiplied by weights $w_{\eta\nu}^{(1)}$
($\nu=1,2,...,k$), shifted by biases $b_{\nu}^{(1)}$, and sent as
input to the first hidden layer ($n=2$). This layer applies a transfer
(activation) function $f^{(2)}(x)$ at each node, producing a set
of outputs $t_{\nu}^{(2)}=f^{(2)}(\sum_{\kappa=1}^{\kappa=m}G_{\kappa}w_{\kappa\nu}^{(1)}+b_{\nu}^{(1)})$.
They are multiplied by another set of weights $w_{\nu\mu}^{(2)}$
($\mu=1,2,...,l$), shifted by biases $b_{\mu}^{(2)}$, and the parameters
$\sum_{\kappa=1}^{\kappa=l}t_{\kappa}^{(2)}w_{\kappa\mu}^{(2)}+b_{\mu}^{(2)}$
are fed into the next layer with $n=3$, which applies to them its
own transfer function $f^{(3)}(x)$, and the process continues. The
last layer ($n=M$) does not change its input parameters $t_{\lambda}^{(M)}$
($\lambda=1,...,m$) and delivers them as the final NN output. The
data flow through the NN can be described by the iteration scheme
\begin{equation}
t_{\eta}^{(n)}=f^{(n)}\left(\sum_{\kappa}t_{\kappa}^{(n-1)}w_{\kappa\eta}^{(n-1)}+b_{\eta}^{(n-1)}\right),\quad n=2,3,...,M,\label{eq:107-1}
\end{equation}
with the initial condition $t_{\eta}^{(1)}=G_{\eta}$ and the final
transfer function $f^{(M)}(x)\equiv x$.

Thus, a feed-forward network implements a nested analytical function
with the weights $w_{\kappa\eta}^{(n)}$ and biases $b_{\eta}^{(n)}$
as fitting parameters. Typical transfer functions are the sigmoidal
function $f(x)=\left(1+e^{-x}\right)^{-1}$ and the hyperbolic tangent
$f(x)=\tanh(x)$, but other functions with similar shapes can also
be used. The reader is referred to Behler's paper \citep{Behler:2011aa}
for an excellent exposition of the NN method, including the exact
count of the total number of parameters and comparison of different
transfer functions.

Although the NN architecture itself has no physical meaning, it provides
hyper-parameters that can be optimized during the training by adding
or removing nodes or whole layers. This is usually accomplished by
trial and error, but evolutionary algorithms and other architecture
search methods have also been proposed. In addition to the feed-forward
architectures, convolutional networks and radial basis function NNs
have been explored, although mostly for organic molecules \citep{Schutt:148aa}.

We emphasize again that in all cases, the regression model is little
more than a black box with many fitting parameters without any physical
significance.

\subsection{The training and validation of ML potentials\label{subsec:Training-ML}}

The regression $\mathcal{R}$ in Eq.(\ref{eq:ML-1}) depends on a
large set of adjustable parameters, which are optimized to best reproduce
the input-output pairs from the DFT dataset. The reference DFT data
comes in the form of energies $\left\{ E_{\textrm{DFT}}^{s}\right\} $
and (often) atomic forces $\left\{ \mathbf{F}_{\textrm{DFT}}^{s}\right\} $
and/or stress tensors $\left\{ \textsf{\ensuremath{\mathbf{T}}}_{\textrm{DFT}}^{s}\right\} $
for a set of supercells. One specific feature of ML potentials, compared
with ML regression problems in other fields, is the absence of one-to-one
correspondence between the input feature vectors and the reference
energies (respectively, forces or stresses). The potential predicts
individual atomic energies, which must be summed over the supercell
atoms before comparing the supercell energy $E^{s}$ with the reference
value $E_{\textrm{DFT}}^{s}$. It should also be noted that some of
the force and stress tensor components can be zero by symmetry and
thus useless for the training process.

The simplest form of the loss function minimized during the potential
training is

\begin{eqnarray}
\mathcal{E} & = & \dfrac{1}{N}\sum_{s=1}^{N}\left(\dfrac{E^{s}-E_{\textrm{DFT}}^{s}}{N_{s}}\right)^{2}+\tau_{1}\dfrac{1}{N}\sum_{s=1}^{N}\sum_{\alpha=1}^{3}\left[F_{\alpha}^{s}-\left(F_{\alpha}^{s}\right)_{\textrm{DFT}}\right]^{2}\nonumber \\
 & + & \tau_{2}\dfrac{1}{N}\sum_{s=1}^{N}\sum_{\alpha,\beta=1}^{3}\left[T_{\alpha\beta}^{s}-\left(T_{\alpha\beta}^{s}\right)_{\textrm{DFT}}\right]^{2}+\tau_{3}\dfrac{1}{L}\sum_{\kappa=1}^{L}\left|p_{\kappa}\right|^{2}.\label{eq:NN3}
\end{eqnarray}
Here, $N_{s}$ is the number of atoms in supercell $s$, $N$ is the
total number of supercells in the database, $L$ is the number of
fitting parameters, and $\tau_{1}$, $\tau_{2}$ and $\tau_{3}$ are
adjustable coefficients treated as hyper-parameters. The first three
terms in the right-hand side represent the mean-square error of fitting.
Some authors write the first term as the mean-square error of the
supercell energy, not the per-atom energy as in Eq.(\ref{eq:NN3}).
The last term in Eq.(\ref{eq:NN3}) is added for regularization purposes
to ensure that the parameters remain sufficiently small for smooth
interpolation. Some ML potentials do not require explicit optimization:
the parameters are obtained by matrix inversion or similar algebraic
operations. In all cases, however, this step must be repeated multiple
times to optimize the hyper-parameters.

Once the final values of the potential parameters are established,
they are fixed and become part of the definition of the ML potential,
just like with the traditional potentials. It is expected that the
potential will make accurate energy and force predictions for new
configurations by interpolating between the DFT points.

Several optimization algorithms are used, the most common of them
being the backpropagation (steepest descent method), the Levenberg-Marquardt,
the Davidon-Fletcher-Powell (DFP), and the Broyden--Fletcher--Goldfarb--Shanno
(BFGS) unconstrained optimization algorithms \citep{Fletcher:1987aa,NRC2}.
The loss function has a rugged terrain with numerous dimples and wrinkles;
hence the gradient methods are easily trapped in local minima. Numerous
restarts from different initial guesses are usually required to reach
a deeper minimum. Some developers start the training with a global
search using, for example, an evolutionary algorithm to avoid the
traps, followed by a gradient-based minimization. Because of the large
size of the reference database, the training is often performed in
a batch mode: the model is optimized on relatively small subsets,
selected from the full database at random or by a chosen rotation
rule until each configuration is exposed multiple times. The precision
of training is measured by either the root-mean-square error (RMSE)
or mean absolute error (MAE).

Many strategies have been developed to avoid underfitting or overfitting
of the database. The overfitting is especially dangerous as it produces
oscillations between the reference points that degrade the predictive
capability of the potential. The common practice is to monitor the
error of predictions on a small validation set excluded from the training
dataset. The divergence between the training and validation errors
signals the onset of overfitting. The $k$-fold cross-validation and
other validation methods are also applied to test the potentials for
overfitting.

Given that the regression models depend on thousands of adjustable
parameters and offer enormous flexibility in fitting the reference
database, it is hardly surprising that most ML potentials readily
achieve the training accuracy of several meV/atom. We emphasize, however,
that this high accuracy is achieved by purely numerical interpolation.
Predicting the energies of new atomic configurations significantly
different from those included in the training database requires extrapolation.
Being a purely numerical procedure, the extrapolation can produce
unpredictable and often meaningless results (Fig.~\ref{fig:Comparison of potentials}b).\footnote{Traditional potentials are often used as an easy target to demonstrate
the superior accuracy of ML potentials. The comparison is unfair as
the two classes of potentials contain drastically different numbers
of fitting parameters ($\mathcal{O}(10)$ and $\mathcal{O}(10^{3})$,
respectively) and are based on different philosophies. One can easily
compile a set of examples of spectacular failures of ML potentials
outside the training domain where a traditional potential makes physically
meaningful (albeit not perfectly accurate) predictions.}

\subsection{The reference database\label{subsec:Database-ML}}

The reference database typically contains $\sim10^{3}$ to $10^{4}$
supercells with energies, forces, and (often) stresses obtained by
routine quantum-mechanical (usually, high-throughput DFT) calculations.
Either static structures or snapshots of AIMD trajectories (or both)
can be included. Alternatively, the structures can be generated by
MD simulations with a current version of the potential, followed by
the energy (force and/or stress) calculations by DFT. While there
are several schools of thought about the most effective procedure
for assembling the database, the consensus is that it should be as
diverse as possible and adequately represent the configurations most
relevant to the intended applications. Preference is usually given
to low-energy configurations, but non-equilibrium structures are also
included to represent transition states and cover a broader domain
in the configuration space.

There are diverging opinions about the role of human expertise in
the ML potential development (as in many other fields of science and
technology). One approach is to judiciously select a set of reference
structures deemed to be most relevant to the targeted applications.
For example, since metals are usually studied for mechanical behavior,
structures representing defects are given preference, especially dislocations,
twin boundaries, stacking faults, and grain boundaries. Several alternate
crystal structures, deformation paths between them, and the liquid
phase are also included to expand the configuration domain. On the
other hand, properties such as the Grueneisen parameter or the phononic
thermal conductivity are considered less important and are rarely
even checked. For a covalent material, such as silicon, the potential
can be expected to reproduce the thermal conductivity, along with
various alternate structures (especially liquid and amorphous) and
defects (especially surfaces and the crack tip). For potentials intended
for a more specific application, the database will be even more specialized.
Assembling such handcrafted datasets requires a fair amount of decision-making
based on the developer's expert knowledge and physical intuition.
The reliance on human expertise is even more significant in developing
binary and multicomponent potentials intended for a broad range of
metallurgical applications \citep{Kobayashi:2017aa,Marchand:2020aa}.
On the downside, a hand-picked database may contain many near-repeat
or uninformative local environments that contribute little to the
accuracy of the final product.

With this approach, the development of an ML potential becomes similar
to that of traditional potentials. In fact, for ML potentials, the
situation is more complicated because the training process reproduces
the PES and not directly the properties. While theoretically, the
PES uniquely defines the properties, in practice, even a tightly fit
PES does not automatically guarantee accurate predictions of properties.
Experience shows that if a set of potentials is trained on the same
database to the same RMSE error (say, 3-5 meV/atom) starting from
different initial conditions, such potentials often predict significantly
different values of the properties. Additional efforts are required
to select a potential with the best combination of properties by testing
multiple versions. This adds a human-controlled feedback loop to the
training process since the notion of a ``best combination'' is difficult
to express by an algorithm. One way to improve a particular property
is to add more reference structures controlling this property. For
some properties, this is straightforward. For example, the inclusion
of additional supercells containing a vacancy or a particular surface
may ensure a more accurate vacancy (respectively, surface) formation
energy. For other properties, such as the melting temperature, the
connection is less direct and more difficult to control.\footnote{One can include additional structures representing the liquid and
the ground-state crystal structures at the expected melting temperature.
In addition to the high computational overhead and poor representation
of liquids by small supercells, this approach cannot guarantee that
the liquid will not crystallize into a wrong crystal structure lying
above the ground state at 0 K but becoming more stable at high temperatures.
It has been shown, however, that DFT melting point calculations can
be accelerated by constructing a ML on-the-fly potential using the
Bayesian inference approach \citep{Jinnouchi:2019aa}.}

Alternatively, algorithms have been proposed for generating the reference
database with little or no human intervention. In most cases, the
database construction and the potential training become part of one
and the same ``active learning'' process \citep{Frederiksen04,Behler_2014,Li:2015aa,Podryabinkin:2017aa,Deringer:2018aa,Bianchini:2019aa,Zhang:2019ab,Bernstein:2019aa,Sivaraman:2020aa,Rosenbrock:2021aa}.
In a typical scenario, a preliminary version of the potential is trained
on an initial DFT database. An MD simulation is performed with this
potential until configurations are encountered that are sufficiently
different from the known ones, signaling that the simulation ran into
a poorly represented region of the configuration space. Different
novelty criteria have been proposed to detect the point where the
simulation drifts outside the reference domain. The ``unknown'' configurations
are then added to the database, and their DFT energies (forces/stresses)
are computed either offline or on-the-fly. The training is continued
on the expanded database, and the updated potential is used to continue
the MD simulation. The iterations are repeated until self-consistency
is reached, i.e., when no new configurations are discovered within
a reasonable MD time. The MD simulation can be replaced by random
structure searches, in which multiple randomized structures are quenched
to find local energy minima. Various other algorithms have been devised
for configuration space exploration and training-testing strategies.

Arguably, the advantage of the active learning approaches is the ability
to automatically generate the most economical reference database for
achieving the desired training accuracy and self-consistent behavior
of the final potential. On the other hand, the configuration domain
covered by the potential depends on the chosen MD simulation protocol
(ensemble, temperatures, densities) or, respectively, on the chosen
algorithm for the generation and testing of the new structures. The
performance of the potential outside this domain remains uncontrollable.
Furthermore, the previous comments about the PES approximation versus
the property predictions are relevant to the present case as well.
Automatically achieving a desired accuracy of training does not automatically
guarantee that the physical property predictions will be accurate
enough for practical applications of the potential. Automation is
possible and makes sense for surrogate models and some special-purpose
potentials. However, a fully automated generation of a general-purpose
potential does not seem to be a reasonable expectation.

\subsection{ML potential software\label{subsec:ML-software}}

Several software packages have been developed for the ML potential
generation and simulations. Some are stand-alone packages, others
are interfaced with the Large-scale Atomic/Molecular Massively Parallel
Simulator (LAMMPS) \citep{Plimpton95}, the Vienna Ab Initio Simulation
Package (VASP) \citep{Kresse1996,Kresse1999}, or other popular software.
In this highly dynamic field, several new packages are released every
year. A few examples are given below in no particular order and without
any claim of completeness.

\emph{ASE} (Atomistic Simulation Environment) \citep{Larsen:2017aa,ASE:aa}
offers a set of Python tools for setting up, manipulating, running,
visualizing, and analyzing atomistic simulations. The environment
is linked to simulation software such as LAMMPS, and to DFT packages
such as VASP and Quantum Espresso \citep{Quantumespresso}. It provides
an excellent platform for DFT database generation and ML potential
testing.

\emph{Amp} (Atomistic Machine-learning Package) \citep{Khorshidi:2016aa,Amp:aa}
is somewhat similar to ASE and additionally contains modules for ML
potential training using an assortment of descriptors and optimization
algorithms. The focus is on NN potentials, but other user-defined
regression models are also accepted. As an example, Amp was applied
to develop a quaternary potential for H and CO on a Mo surface \citep{Khorshidi:2016aa}.

\emph{N2P2} (Neural Network Potential Package) \citep{N2P2} provides
a repository and tools for the training of Behler-Parrinello type
\citep{Behler07} NN potentials and calculation of energies and forces
using such potentials. It also contains components needed for running
MD simulations with LAMMPS.

\emph{Aenet} (Atomic Energy Network) \citep{Artrith:2016aa,Artrith:2017aa,Cooper:2020aa,Aenet}.
Software package for the training and usage of Behler-Parrinello type
\citep{Behler07} NN potentials. Modules for the energy and force
calculations with the potentials are also included.

\emph{MLIP} (Machine Learning Interatomic Potentials) \citep{Novikov:2021aa,MLIP}
package constructs MTP potentials \citep{Shapeev:2016aa,Podryabinkin:2017aa,Gubaev:2019aa,Novikov:2021aa}
(including multicomponent potentials \citep{Gubaev:2019aa}) by the
active learning approach.

\emph{KLIFF} (KIM-based Learning-Integrated Fitting Framework) is
a package intended for the development of both traditional and NN
interatomic potentials. The package is linked to the OpenKIM project
\citep{OpenKim}, and through it, to a large interatomic potential
repository and LAMMPS simulations.

\emph{MAISE} (Module for ab initio structure evolution) \citep{Hajinazar:2021aa}
is a package for automated generation of NN (Behler-Parrinello type
\citep{Behler07}) potentials for global structure optimization. The
DFT database is generated automatically by an evolutionary structure
sampling procedure.

\subsection{Discussion of ML potentials\label{subsec:Discussion-of-ML}}

The distinguishing features of the ML potentials compared with the
traditional potentials are summarized in Table \ref{tab:Comparison}.
When properly trained, an ML potential can predict the energy and
forces with nearly DFT accuracy, provided the atomic configuration
bears enough similarity with some of the known configurations from
the training database. The high accuracy is achieved by numerical
interpolation using a high-dimensional regression model trained on
a massive DFT database. The regression maps the local atomic environments,
encoded in local structural parameters (fingerprints), onto the PES
of the material. By contrast to the traditional potentials, which
aim to reproduce a particular set of physical properties, ML potentials
approximate the PES (actually, only some part of it), with the expectation
that accurate values of physical properties will follow. Another crucial
difference is that the ML potentials are not based on any physical
considerations. The fingerprints and the regression only respect the
locality of atomic interactions and the invariance of energy under
translations, rotations, and relabeling of atoms but are otherwise
devoid of any physical significance. Accordingly, the ML potential
are sometimes called ``mathematical'' or ``non-physical'' \citep{Behler:2016aa}.

Freedom from physics is a double-edged sword. ML potentials are not
specific to any particular class of materials. A potential for almost
any material can be developed with the same model, same software,
and a comparable amount of computational effort regardless of the
type of chemical bonding. On the other hand, ML potentials are powerful
numerical interpolators but poor extrapolators. The energy/force predictions
for less familiar atomic environments are based on a purely mathematical
extrapolation procedure whose results are unpredictable and often
physically meaningless. Special efforts are required to keep the simulations
within or close to the configuration domain on which the potential
was trained.

Given the limited transferability, there are two approaches to utilizing
the ML potentials. One is to develop a special-purpose potential for
a particular task without claiming any broader applicability. The
approach exploits the ML potentials' high accuracy and computational
efficiency (compared to DFT) while keeping the simulation in the interpolation
regime. One recent example is developing a GAP Si potential specifically
designed for phonon properties and thermal conductivity calculations,
including the effect of vacancies \citep{Babaei:2019aa}. Similarly,
a deep NN potential was constructed to calculate the vacancy formation
free energy in Al as a function of temperature \citep{Bochkarev_2019}.
Another excellent example of using a special-purpose potential (deep
NN in this case) is the recent study of the nucleation of crystalline
Si from the liquid phase \citep{Bonati:2018aa}.

The second approach is to assemble a large and highly diverse reference
database covering as broad a configuration domain as possible, including
atomic environments that (1) typically occur in atomistic simulations,
and (2) are most relevant to the set of physical properties expected
to be reproduced by potentials. Simulations with a potential trained
on this database will then occur predominantly in the interpolation
mode. Theoretically, there is still a chance that the simulation will
wander away to unexplored territory or will be trapped in a gap (dark
pocket) inside the reference domain. Nevertheless, some of the recent
potentials developed with this approach, such as the GAP potentials
for Si \citep{Bartok_2018} and C \citep{Rowe:2020aa}, do reproduce
an incredibly broad spectrum of physical properties with high accuracy.
Based on the broad applicability, the developers place these potentials
in the general-purpose category.

Another usage of ML potentials is to provide surrogate models to accelerate
DFT calculations. This approach has been especially effective in the
area of crystal structure searches \citep{Deringer:2018ab,Ibarra-Hernandez:2018aa,Oganov:2019aa,Hajinazar:2019aa,Podryabinkin:2019aa}
combined with active learning by either structure relaxations \citep{Podryabinkin:2019aa,Gubaev:2019aa}
or evolutionary algorithms \citep{Bisbo:2020aa,Hajinazar:2021aa}.
The potential is constructed as part of the structure search and accelerates
the search by orders of magnitude compared to high-throughput DFT
calculations. In some cases, stable binary or ternary structures have
been revealed that were missed by DFT searches. Acceleration of DFT
calculations by on-the-fly constructed ML potentials has been proposed
for several other applications, such as the melting temperature calculations
\citep{Jinnouchi:2019aa}. In all such cases, the ML potential only
plays the role of a temporary construction not intended for independent
applications.

\section{Physically-informed machine-learning potentials\label{sec:Physically-informed-ML}}

\begin{figure}[h]
\noindent \begin{centering}
\includegraphics[width=0.5\textwidth]{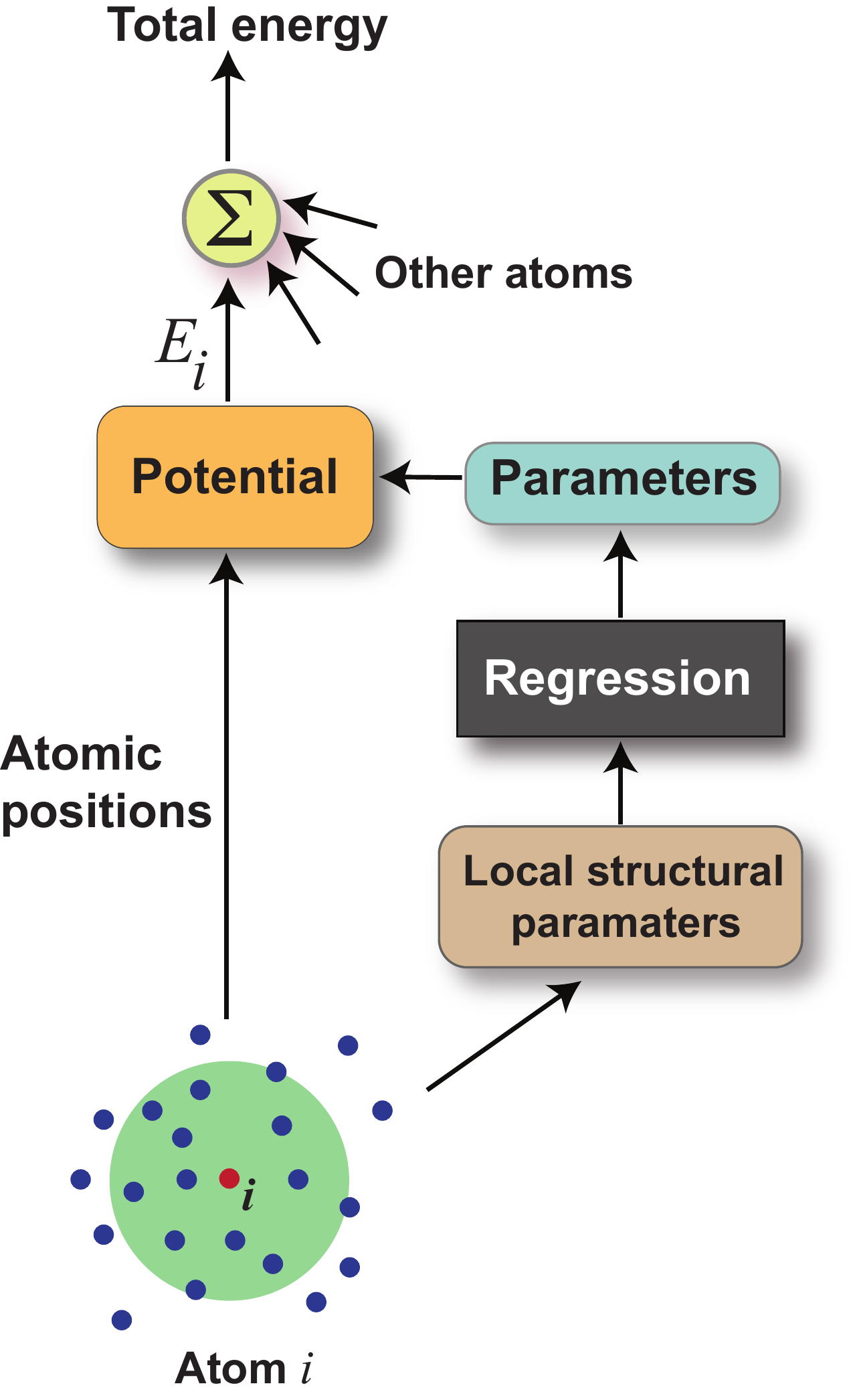}
\par\end{centering}
\caption{Flowchart of the total energy calculation with a physically-informed
ML interatomic potential. The local environment of an atom $i$ within
the cutoff sphere (green) is encoded in a set of local structural
parameters, which are then mapped onto a set of parameters of a physics-based
interatomic potential. These parameters and the local atomic coordinates
are used to compute the energy $E_{i}$ assigned to atom $i$. The
summation of the energies of other atoms of the system (symbol $\Sigma$)
gives the total energy and thus a point on the PES of the system.\label{fig:Flowchart-PINN}}
\end{figure}

As discussed in the previous sections, many of the strengths and weaknesses
of the traditional and ML potentials are complementary to each other
(Table \ref{tab:Comparison}). A new direction has recently emerged,
aiming to strike a golden compromise between the two by taking the
best from both worlds. This goal can be achieved by choosing a general
enough form on a physics-based interatomic potential and letting a
ML regression predict its parameters according to each atom's local
environment. In contrast to Eq.(\ref{eq:ML-1}) describing a mathematical
ML potential, the formula of a \emph{physically-informed ML potential}
is 
\begin{equation}
\mathbf{R}_{i}\rightarrow\mathbf{G}_{i}\overset{\mathcal{R}}{\rightarrow}\mathbf{p}_{i}\overset{\Phi}{\rightarrow}E_{i},\label{eq:ML-2}
\end{equation}
see the flowchart in Fig.~\ref{fig:Flowchart-PINN}. Instead of directly
predicting the atomic energy $E_{i}$, the regression $\mathcal{R}$
outputs a set of potential parameters $\mathbf{p}_{i}$ most appropriate
for the environment of the particular atom $i$. The atomic energy
$E_{i}$ is then computed with the interatomic potential $\Phi(\mathbf{R}_{i},\mathbf{p}_{i})$.
In other words, the regression output is ``piped'' through a model
whose functional form ensures that the energy predictions are physically
meaningful. Extrapolation to new environments is now guided by the
physics embodied in the interatomic potential rather than a purely
mathematical procedure.

Any model eventually fails. Physically-informed ML potentials can
also produce unphysical results when taken too far away from the familiar
territory. However, the physics-guided extrapolation is likely to
expand the potential's reliability domain compared with purely mathematical
models (Fig.~\ref{fig:Comparison of potentials}c).

This general idea can be realized with any regression method and any
suitable interatomic potential. The recently developed physically-informed
neural network (PINN) method \citep{PINN-1} relies on a NN regression
and an analytical bond-order potential (BOP) whose functional form
is general enough to apply to both metals and nonmetals. A detailed
description of the BOP potential can be found in Appendix B. In short,
the potential captures pairwise repulsion and attraction of atoms,
the bond-order effect (bond weakening with the number of bonds), angular
dependence of the bond energies, screening of bonds by surrounding
atoms, and the promotion energy. The interactions are limited to a
coordination sphere with a smooth cutoff. The local atomic environments
are represented by Gaussian descriptors specified in Appendix C, but
we emphasize that other choices of the descriptors are equally possible.

\begin{figure}
\includegraphics[width=1\textwidth]{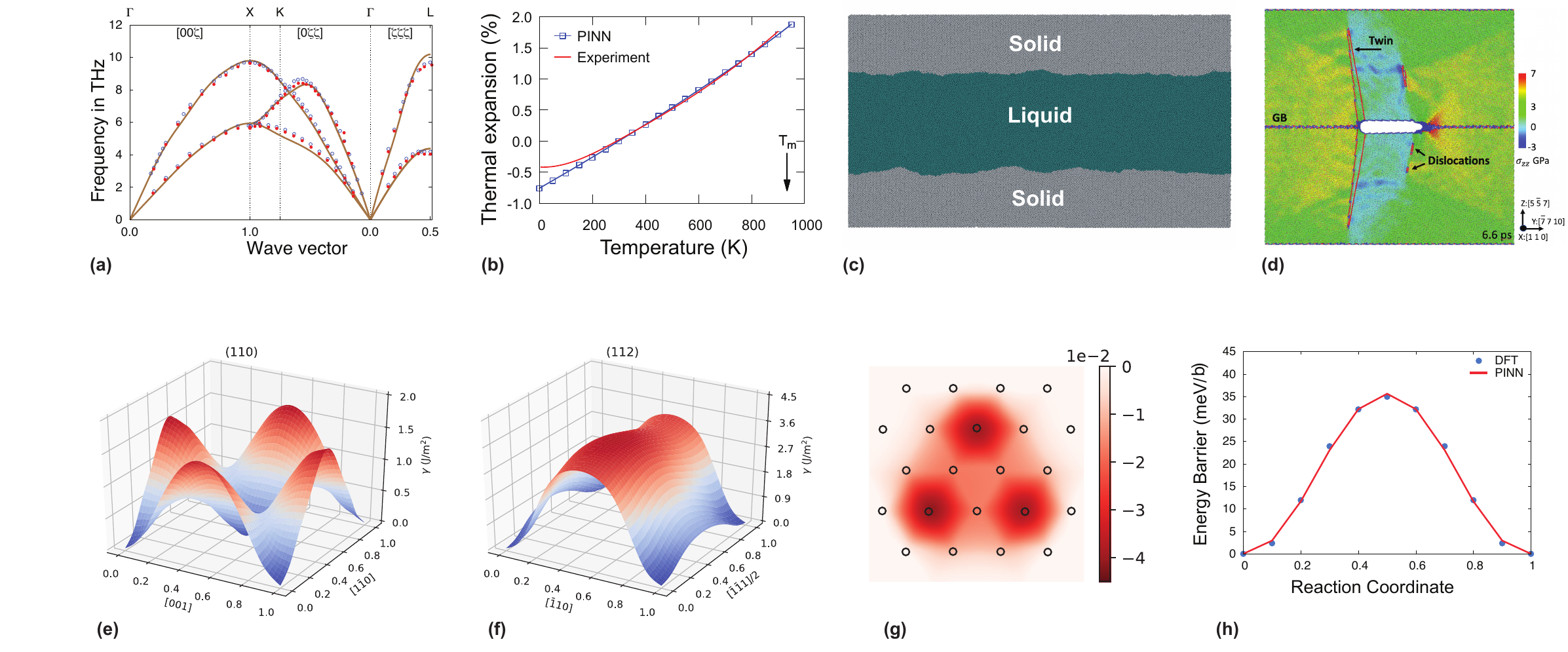}\caption{Examples of properties calculated with the PINN potentials for (a-d)
Al \citep{Pun:2020aa} and (e-h) Ta \citep{Ta-PINN-in-review}. (a)
Phonon dispersion relations computed with the PINN potential (curves)
compared with experimental data (points). (b) Linear thermal expansion
relative to room temperature predicted by the PINN potential (points)
compared with experiment (curve). (c) Simulation block used for computing
the solid-liquid interface tension by the capillary fluctuation method.
(d) MD simulation of crack nucleation and growth on a grain boundary.
(e,f) $\gamma$-surfaces in body-centered cubic Ta on (110) and (112)
planes, respectively. (g) Nye tensor plot of the core structure of
the $\frac{1}{2}\left\langle 111\right\rangle $ screw dislocation
in Ta predicted by the PINN potential.\textbf{ }(h) Peierls barrier
of the $\frac{1}{2}\left\langle 111\right\rangle $ screw dislocation
predicted by the PINN potential (lines) in comparison with DFT calculations
(points).\label{fig:Examples-Al-Ta}}
\end{figure}

The original PINN formulation \citep{PINN-1} was recently improved
\citep{Pun:2020aa,Ta-PINN-in-review} by introducing a global version
of the BOP potential trained on the entire reference database. After
the training, the optimized parameters $\mathbf{p}^{0}=(p_{i1}^{0},...,p_{im}^{0})$
are fixed and become part of the potential definition. Since this
parameter set is relatively small ($m\sim10$), the error of fitting
is usually on the order of $10^{2}$ meV/atom. The role of the NN
is to add to $\mathbf{p}^{0}$ a set of local ``perturbations'' $\delta\mathbf{p}_{i}=(\delta p_{i1},...,\delta p_{im})$.
The final parameter set $\mathbf{p}_{i}=\mathbf{p}^{0}+\delta\mathbf{p}_{i}$
is then used to predict the atomic energy $E_{i}=\Phi(\mathbf{R}_{i},\mathbf{p}^{0}+\delta\mathbf{p}_{i})$.
The magnitudes of the perturbations are kept as small as possible.
Their goal is to achieve the DFT level of accuracy of the training.
In this model, the energy predictions are largely guided by the global
BOP potential $\Phi(\mathbf{R}_{i},\mathbf{p}^{0})$ ensuring a smooth
and physically meaningful extrapolation outside the training domain.
This scheme has been shown \citep{Pun:2020aa,Ta-PINN-in-review} to
significantly improve the transferability without compromising the
accuracy or increasing the computational cost compared with the original
formulation \citep{PINN-1}.

General-purpose PINN potentials have been constructed for Al \citep{PINN-1,Pun:2020aa}
and Ta \citep{Ta-PINN-in-review}, with several other potentials being
currently developed. Both the Al and Ta potentials accurately describe
a broad spectrum of properties of these metals, including the mechanical
and thermal properties most relevant to materials science applications.
A select set of examples is given in Fig.~\ref{fig:Examples-Al-Ta}.
A multicomponent version of PINN has been formulated and tested, and
is currently being used to develop binary potentials. The computational
efficiency of the PINN Al potential has been evaluated \citep{Pun:2020aa}.
Although the specific numbers depend on the computational software,
hardware, and the type of tests, as a general guide, the PINN potential
is two orders of magnitude slower than the EAM Al potential \citep{Mishin99b}.
The computational overhead due to the BOP potential is about 25 \%
of the total time. Considering that ML potentials are orders of magnitude
faster than straight DFT calculations, this modest overhead can be
considered a small price for the improved transferability.

The general idea of incorporating physics into ML models was explored
by several authors in the past. Skinner and Broughton \citep{Skinner_1995}
demonstrated that artificial NNs can be used to construct Lennard-Jones
and Stilliger-Weber potentials. Other authors developed ReaxFF, BOP,
and other traditional potentials by generating a massive DFT database
and applying training algorithms adapted from the ML field \citep{Cherukara:2016vk,Chan:2019ta}.
Drautz's ACE potentials \citep{Drautz:2019aa,Drautz:2020tl} are based
on a physics-motivated and physically interpretable cluster expansion
containing a set of adjustable parameters. While we classify the ACE
potentials as traditional, they include some features of ML potentials,
such as local structural descriptors and the parameter optimization
on a large DFT database. On the other hand, there is a recent trend
to include physics-inspired terms in mathematical ML potentials. Glielmo
et al.~\citep{Glielmo:2018aa} described a construction of $n$-body
Gaussian process kernels that capture the $n$-body nature of atomic
interactions in physical systems. Additional analytical terms mimicking
either short-range pairwise repulsion or long-range van der Waals
interactions are included in recent versions of GAP potentials \citep{Bartok_2018,Rowe:2020aa}.
The parameters describing such terms are fitted separately from the
GAP part and then subtracted from the total energy during the training.
Perhaps the closest to the PINN model was the work by Malshe et al.~\citep{Malshe:2008aa},
who constructed a Tersoff potential for the Si$_{5}$ cluster in which
the potential parameters were controlled by a pre-trained NN. In their
model, the potential parameters were not fixed but varied in the course
of MD simulations according to the instantaneous atomic positions.

The physically-informed ML potentials, of which PINN one an example,
do not simply fill the gap between the classical and ML potentials
but build upon both to create a new and distinct class of interatomic
potentials. Like traditional potentials, they adopt an atomic interaction
model explicitly describing diverse physical effects, such as the
many-body character of interactions, the bond-order effect, and the
bond screening by neighbors. (In the future, more effects can be added,
such as magnetism.) By contrast to the traditional potentials, the
parameters of the physically-informed ML potentials are dynamic: they
are locally adjusted during the simulation in response to the ever-changing
local atomic environments. At the same time, potentials of this class
have the descriptor-regressor structure shared by all ML potentials
and are trained on a DFT database using the statistical learning methods.

\section{Summary and outlook\label{sec:Summary}}

ML potentials have emerged as a powerful new tool for materials modeling
and new materials discovery. When used within the boundaries of validity,
an ML potential can predict the energy and atomic forces with nearly
DFT accuracy but orders of magnitude faster. The computational time
scales linearly with the number of atoms $N$, easily beating the
computational efficiency of the $N^{3}$-scaled DFT calculations for
large systems. This enables researchers to extend DFT-level calculations
to much larger systems and much longer MD simulation times. The development
of ML potentials is leveraged by the availability of massive DFT databases
generated by high-throughput calculations. The accuracy of the ML
potentials can be improved systematically by augmenting the reference
database with new structures and continuing the training process.
The flexibility of the ML potentials is enormous. The same potential
format and the same training algorithm can be applied to different
classes of materials regardless of the nature of the chemical bonding.

Like any model, ML potentials have their limitations. In our view,
the major limitation is the lack of physics-based transferability
to unknown structures. Other than smoothness, locality, and invariance
of energy, no physics specific to the particular system is included
in the ML potential. Being purely mathematical constructions, ML potentials
are little more than accurate numerical interpolators of DFT databases.
Predictions of physical properties outside the interpolation domain
are based on a mathematical algorithm and can give uncontrollable
and often physically meaningless results. The risk of generating inaccurate
predictions can be mitigated by monitoring the simulation process
to detect departures of the trajectory from the training domain. Another
approach is to develop the potential concurrently with the DFT database
construction (e.g., by on-the-fly training). The system will then
likely remain within the training domain as long as the simulation
remains similar to those used during the training.

The only way to ensure that the extrapolation to unknown configurations
makes physical sense is to inform the model about some basic rules
of physics. This approach is pursued by the proposed physically-informed
ML potentials. The integration of an ML regression with a physics-based
interatomic potential preserves the DFT accuracy of training without
increasing the number of fitting parameters or paying any significant
computational overhead. At the same time, the transferability is improved
compared with purely mathematical models, opening the door for the
design of general-purpose ML potentials. The NN-based PINN potentials
demonstrate a promise of this approach\footnote{The improved transferability of PINN potentials has been demonstrated
by comparison with NN potentials \citep{PINN-1} and other purely
mathematical ML models (unpublished). Comparison with the recent GAP
potentials containing physically-inspired analytical terms \citep{Bartok_2018,Rowe:2020aa}
will require additional work in the future.} and could serve as a springboard for developing similar models with
other choices of the regression model, descriptors, or the physics-based
potential.

Many materials science applications require the development of reliable
potentials for alloy systems. The ability of ML potentials to reproduce
alloy properties across wide composition ranges, including phase diagram
calculations, has been recently demonstrated \citep{Nyshadham:2019aa,Rosenbrock:2021aa}.
It should be noted that the application of both traditional and ML
potentials to multicomponent systems is not as straightforward as
in DFT calculations. Each time we add a new chemical element to the
system, a new potential must be constructed, which is a demanding
task. ML potentials overcome the problem of incompatible functional
forms between the metallic and nonmetallic elements inherent in the
traditional potentials.

On the other hand, most of the traditional potentials are ``inheritable''.
Existing elemental potentials can be included into a binary potential
by only fitting the cross-interaction parameters; the binary potential
can be then crossed with another elemental potential to obtain a ternary
potential, and so on. This strategy saves efforts and enables accurate
comparison of alloy systems formed by the same base element with different
solutes. Most of the ML potentials are incapable of inheriting elemental
potentials. This unfortunate feature may result in a proliferation
of different potentials for the same element developed as a stand-alone
potential or as part of binaries or ternaries. This is not a severe
problem for the ``make it and forget it'' potentials, such as the
surrogate models or some of the special-purpose potentials constructed
for just one particular simulation. However, for broadly applicable
potentials intended for multi-purpose utilization by many groups,
the inheritance is a highly desirable feature. For NN potentials,
a stratified construction procedure was proposed \citep{Hajinazar:2017aa}
and successfully applied to the evolutionary structure searches in
multicomponent systems. In the future, generating accurate general-purpose
potentials by this or similar procedures could be pursued. It is worth
exploring if other forms of ML potentials could also be modified to
enable the inheritance feature.

When the traditional interatomic potentials first appeared in the
1980s, initially the main focus was on demonstrating their capabilities.
The initial excitement about their ability to make quantitative predictions
was followed by recognizing their limitations and a better understanding
of the application domain. In the mid-1990s, the field entered a new
phase in which the focus shifted toward practical applications. The
method development still continued: many new potentials were constructed,
several simulation packages were developed, and repositories of existing
potentials began to grow. However, the main goal of the atomistic
simulations became to gain some \emph{new knowledge} about the materials.
The following question could now be asked and often answered: \emph{What
have we learned about the material from this simulation that we did
not know previously?} Eventually, the potential-based simulations
took their proper place among other computational tools operating
on different length and time scales.

The ML potential field is likely to undergo a similar evolution. It
is currently on the rising branch of the hype curve. The publication
rate is rising rapidly as new groups are drawn into the field by the
high promise of the ML potentials and/or fascination by all things
ML. The overwhelming majority of the publications are concentrated
on the method development: demonstration of the new capabilities,
the search for more effective descriptors and regression models, design
of new algorithms to optimize and automatize the DFT database construction,
and computational benchmarking. Most of the ML potentials published
so far are proof-of-principle type, rarely used by other groups after
the publication. In most cases, the success is measured by the ability
of the potential to reproduce already known (although often complex)
structures or properties. While very important for the method development,
these efforts \emph{per se} do not generate new knowledge of the materials.

The overexcitement will eventually subside, and ML potentials will
become an integral part of the standard toolkit for materials modeling
alongside other methods. The focus will gradually shift toward answering
the ``What have we learned about the material'' question. Recent
years have already seen several applications of ML potentials that
begin to generate new knowledge of physics and/or materials \citep{Deringer:2019aa}.
Such applications typically rely on special-purpose potentials, often
created as surrogate models interpolating a DFT database. A GAP potential
for the phase-change Ge-Sb-Te material \citep{Mocanu:2018aa} was
used to generate an ensemble of representative glass structures, which
helped understand the electronic nature of mid-gap defect states in
memory materials \citep{Konstantinou:2019aa}. GAP potentials have
provided new insights into the energetics and structures of the numerous
hypothetical polymorphs of carbon, boron, and phosphorous \citep{Deringer:2017aa,Deringer:2018aa,Deringer:2018ab,Podryabinkin:2019aa,Bernstein:2019aa,Rowe:2020aa}.
Caro et al.~\citep{Caro:2018aa} applied a GAP potential to elucidate
the mechanisms of amorphous carbon growth \citep{Caro:2018aa}. A
deep NN Si potential combined with meta-dynamics helped understand
the early stages of nucleation during Si crystallization from the
melt \citep{Bonati:2018aa}. We especially emphasize the recent applications
of ML potentials to the modeling of metallurgical processes, such
as precipitation hardening in Al-based alloys \citep{Kobayashi:2017aa,Marchand:2020aa}
and the deformation behavior of magnesium alloys \citep{Stricker:2020aa}.
The leading thread of these papers is still the demonstration of capabilities
of ML potentials compared with traditional potentials. However, these
papers turn the ML potential field toward the core problems of classical
metallurgy.

We envision that, within the next several years, most of the methodology
will be established, and the field will enter the second phase focused
on discovering new and/or explaining known materials phenomena or
predicting properties that cannot be computed otherwise. We also believe
that the field will turn around to physics by integrating the remarkable
flexibility and adaptivity of the mathematical ML potentials with
tighter guidance from physical models.

\bigskip{}

\noindent \textbf{Acknowledgements}

\noindent I am grateful to Ian Chesser, James Hickman, Raj Koju, Ganga
Purja Pun and Vesselin Yamakov for reading the manuscript and providing
helpful feedback. This work was supported by the Office of Naval Research
under Award No.~N00014-18-1-2612.


\begin{thebibliography}{100}

\bibitem{HMM}
S.~Yip (Editor), \emph{Handbook of Materials Modeling} (Springer, Dordrecht,
  The Netherlands, 2005).

\bibitem{Hafner00}
J.~Hafner, Atomic-scale computational materials science, Acta Mater.
  \textbf{48}, 71--92 (2000).

\bibitem{Giessen:2020aa}
E.~van~der Giessen, P.~A. Schultz, N.~Bertin, V.~V. Bulatov, W.~Cai,
  G.~Cs{\'a}nyi, S.~M. Foiles, M.~G.~D. Geers, C.~Gonz{\'a}lez, M.~H{\"u}tter,
  W.~K. Kim, D.~M. Kochmann, J.~LLorca, A.~E. Mattsson, J.~Rottler, A.~Shluger,
  R.~B. Sills, I.~Steinbach, A.~Strachan and E.~B. Tadmor, Roadmap on
  multiscale materials modeling, Model. Simul. Mater. Sci. Eng. \textbf{28},
  043001 (2020).

\bibitem{Daw83}
M.~S. Daw and M.~I. Baskes, Semiempirical, quantum mechanical calculation of
  hydrogen embrittlement in metals, Phys. Rev. Lett. \textbf{50}, 1285--1288
  (1983).

\bibitem{Daw84}
M.~S. Daw and M.~I. Baskes, Embedded-atom method: Derivation and application to
  impurities, surfaces, and other defects in metals, Phys. Rev. {\rm B}
  \textbf{29}, 6443--6453 (1984).

\bibitem{Tersoff88}
J.~Tersoff, New empirical approach for the structure and energy of covalent
  systems, Phys. Rev. {\rm B} \textbf{37}, 6991--7000 (1988).

\bibitem{Tersoff:1988dn}
J.~Tersoff, Empirical interatomic potential for silicon with improved elastic
  properties, Phys. Rev. {\rm B} \textbf{38}, 9902--9905 (1988).

\bibitem{Tersoff:1989wj}
J.~Tersoff, Modeling solid-state chemistry: Interatomic potentials for
  multicomponent systems, Phys. Rev. {\rm B} \textbf{39}, 5566--5568 (1989).

\bibitem{Stillinger85}
F.~H. Stillinger and T.~A. Weber, Computer simulation of local order in
  condensed phases of silicon, Phys. Rev. {\rm B} \textbf{31}, 5262--5271
  (1985).

\bibitem{Brenner00}
D.~W. Brenner, The art and science of an analytical potential, Phys. Stat.
  Solidi {\rm (b)} \textbf{217}, 23--40 (2000).

\bibitem{Mishin.HMM}
Y.~Mishin, Interatomic potentials for metals, in \emph{Handbook of Materials
  Modeling}, edited by S.~Yip, chapter 2.2, pages 459--478 (Springer,
  Dordrecht, The Netherlands, 2005).

\bibitem{Becker:2013aa}
C.~A. Becker, F.~Tavazza, Z.~T. Trautt and R.~A. {Buarque de Macedo},
  Considerations for choosing and using force fields and interatomic potentials
  in materials science and engineering, Current Opinion in Solid State and
  Materials Science \textbf{17}, 277 -- 283 (2013).

\bibitem{Hale_2018}
L.~M. Hale, Z.~T. Trautt and C.~A. Becker, Evaluating variability with
  atomistic simulations: the effect of potential and calculation methodology on
  the modeling of lattice and elastic constants, Modelling and Simulation in
  Materials Science and Engineering \textbf{26}, 055003 (2018).

\bibitem{NIST-Potentials}
\mbox{NIST Interatomic Potentials Repository}:
  http://www.ctcms.nist.gov/potentials/ (Website DOI: 10.18434/m37).

\bibitem{Tadmor:2013aa}
E.~B. Tadmor, R.~S. Elliott, S.~R. Phillpot and S.~B. Sinnott, {NSF}
  cyberinfrastructures: {A} new paradigm for advancing materials simulations,
  Current Opinion in Solid State and Materials Science \textbf{17}, 298--304
  (2013).

\bibitem{OpenKim}
{Knowledgebase of Interatomic Models:} https://openkim.org (OpenKim).

\bibitem{Brommer07}
P.~Brommer and F.~Gahler, Potfit: Effective potentials from ab-initio data,
  Model. Simul. Mater. Sci. Eng. \textbf{15}, 295--304 (2007).

\bibitem{Potfit-2015}
P.~Brommer, A.~Kiselev, D.~Schopf, P.~Beck, J.~Roth and H.-R. Trebin, Classical
  interaction potentials for diverse materials from ab initio data: {A }review
  of potfit, Model. Simul. Mater. Sci. Eng. \textbf{23}, 074002 (2015).

\bibitem{Potfit}
 (Potfit Website: https://www.potfit.net/wiki/doku.php?id=start).

\bibitem{KLIFF}
 (KLIFF Website: https://kliff.readthedocs.io/en/latest/index.html).

\bibitem{Stukowski:2017ui}
A.~Stukowski, E.~Fransson, M.~Mock and P.~Erhart, Atomicrex---a general purpose
  tool for the construction of atomic interaction models, Model. Simul. Mater.
  Sci. Eng. \textbf{25}, 055003 (2017).

\bibitem{Behler:2016aa}
J.~Behler, Perspective: Machine learning potentials for atomistic simulations,
  Phys. Chem. Chem. Phys. \textbf{145}, 170901 (2016).

\bibitem{Botu:2017aa}
V.~Botu, R.~Batra, J.~Chapman and R.~Ramprasad, Machine learning force fields:
  Construction, validation, and outlook, The Journal of Physical Chemistry C
  \textbf{121}, 511--522 (2017).

\bibitem{Deringer:2019aa}
V.~L. Deringer, M.~A. Caro and G.~Cs{\'a}nyi, Machine learning interatomic
  potentials as emerging tools for materials science, Advanced Materials
  \textbf{31}, 1902765 (2019).

\bibitem{Zuo:2020aa}
Y.~Zuo, C.~Chen, X.~Li, Z.~Deng, Y.~Chen, J.~Behler, G.~Cs{\'a}nyi, A.~V.
  Shapeev, A.~P. Thompson, M.~A. Wood and S.~P. Ong, Performance and cost
  assessment of machine learning interatomic potentials, The Journal of
  Physical Chemistry A \textbf{124}, 731--745 (2020).

\bibitem{Morawietz:2020aa}
T.~Morawietz and N.~Artrith, Machine learning-accelerated quantum
  mechanics-based atomistic simulations for industrial applications, Journal of
  Computer-Aided Molecular Design  (2020).

\bibitem{Mueller:2020vy}
T.~Mueller, A.~Hernandez and C.~Wang, Machine learning for interatomic
  potential models, The Journal of Chemical Physics \textbf{152}, 050902
  (2020).

\bibitem{Skinner_1995}
A.~J. Skinner and J.~Q. Broughton, Neural networks in computational materials
  science: training algorithms, Modelling and Simulation in Materials Science
  and Engineering \textbf{3}, 371--390 (1995).

\bibitem{Blank:1995aa}
T.~B. Blank, S.~D. Brown, A.~W. Calhoun and D.~J. Doren, Neural network models
  of potential energy surfaces, J. Chem. Phys. \textbf{103}, 4129--4137 (1995).

\bibitem{Raff:2012aa}
L.~M. Raff, R.~Komanduri, M.~Hagan and S.~T.~S. Bukkapatnam, \emph{Neural
  networks in chemical reaction dynamics} (Oxford University Press, New York,
  NY, 2012).

\bibitem{Mueller:2016aa}
T.~Mueller, A.~G. Kusne and R.~Ramprasad, Machine learning in materials
  science: {Recent} progress and emerging applications, in \emph{Reviews in
  Computational Chemistry}, edited by A.~L. Parrill and K.~B. Lipkowitz,
  volume~29, chapter~4, pages 186--273 (Wiley, 2016).

\bibitem{Ramprasad:2017aa}
R.~Ramprasad, R.~Batra, G.~Pilania, A.~Mannodi-Kanakkithodi and C.~Kim, Machine
  learning in materials informatics: recent applications and prospects, npj
  Computational Materials \textbf{3}, 54 (2017).

\bibitem{Gubernatis:2018aa}
J.~E. Gubernatis and T.~Lookman, Machine learning in materials design and
  discovery: Examples from the present and suggestions for the future, Phys.
  Rev. Materials \textbf{2}, 120301 (2018).

\bibitem{Rickman:2019aa}
J.~Rickman, T.~Lookman and S.~Kalinin, Materials informatics: From the
  atomic-level to the continuum, Acta Materialia \textbf{168}, 473 -- 510
  (2019).

\bibitem{Ong:2019aa}
S.~P. Ong, Accelerating materials science with high-throughput computations and
  machine learning, Computational Materials Science \textbf{161}, 143 -- 150
  (2019).

\bibitem{Pablo:2019aa}
J.~J. de~Pablo, N.~E. Jackson, M.~A. Webb, L.-Q. Chen, J.~E. Moore, D.~Morgan,
  R.~Jacobs, T.~Pollock, D.~G. Schlom, E.~S. Toberer, J.~Analytis, I.~Dabo,
  D.~M. DeLongchamp, G.~A. Fiete, G.~M. Grason, G.~Hautier, Y.~Mo, K.~Rajan,
  E.~J. Reed, E.~Rodriguez, V.~Stevanovic, J.~Suntivich, K.~Thornton and J.-C.
  Zhao, New frontiers for the materials genome initiative, npj Computational
  Materials \textbf{5}, 41 (2019).

\bibitem{Picklum:2019aa}
M.~Picklum and M.~Beetz, Matcalo: Knowledge-enabled machine learning in
  materials science, Computational Materials Science \textbf{163}, 50 -- 62
  (2019).

\bibitem{DeCost_2020}
B.~DeCost, J.~Hattrick-Simpers, Z.~Trautt, A.~Kusne, E.~Campo and M.~Green,
  Scientific {AI} in materials science: a path to a sustainable and scalable
  paradigm, Machine Learning: Science and Technology \textbf{1}, 033001 (2020).

\bibitem{Saal:2020aa}
J.~E. Saal, A.~O. Oliynyk and B.~Meredig, Machine learning in materials
  discovery: Confirmed predictions and their underlying approaches, Annual
  Review of Materials Research \textbf{50}, 49--69 (2020).

\bibitem{Batra:2020aa}
R.~Batra, L.~Song and R.~Ramprasad, Emerging materials intelligence ecosystems
  propelled by machine learning, Nature Reviews Materials  (2020).

\bibitem{Bereau:2015}
T.~Bereau, D.~Andrienko and O.~A. {von Lilienfeld}, Transferable atomic
  multipole machine learning models for small organic molecules, J. Chem.
  Theor. Comput. \textbf{11}, 3225--3233 (2015).

\bibitem{Schutt:148aa}
K.~T. Schutt, H.~E. Sauceda, P.~J. Kindermans, A.~Tkatchenko and K.~R. Muller,
  Schnet - a deep learning architecture for molecules and materials, J. Chem.
  Phys. \textbf{2018}, 241722 (148).

\bibitem{Bereau:2018aa}
T.~Bereau, R.~A. {DiStasio}, A.~Tkatchenko and O.~A. {von Lilienfeld},
  Non-covalent interactions across organic and biological subsets of chemical
  space: {Physics-based potentials} parametrized from machine learning, J.
  Chem. Phys. \textbf{148}, 241706 (2018).

\bibitem{Cooper:2020aa}
A.~M. Cooper, J.~K{\"a}stner, A.~Urban and N.~Artrith, Efficient training of
  {ANN} potentials by including atomic forces via {Taylor} expansion and
  application to water and a transition-metal oxide, npj Computational
  Materials \textbf{6}, 54 (2020).

\bibitem{hohenberg64:dft}
P.~Hohenberg and W.~Kohn, Inhomogeneous electron gas, Phys. Rev. \textbf{136},
  B864--B871 (1964).

\bibitem{kohn65:inhom_elec}
W.~Kohn and L.~J. Sham, Self-consistent equations including exchange and
  correlation effects, Phys. Rev. \textbf{140}, A1133--A1138 (1965).

\bibitem{Curtarolo:2012qf}
S.~Curtarolo, W.~Setyawan, S.~Wang, J.~Xue, K.~Yang, R.~H. Taylor, L.~J.
  Nelson, G.~L.~W. Hart, S.~Sanvito, M.~Buongiorno-Nardelli, N.~Mingo and
  O.~Levy, Aflowlib.org: A distributed materials properties repository from
  high-throughput ab initio calculations, Computational Materials Science
  \textbf{58}, 227--235 (2012).

\bibitem{Curtarolo:2013aa}
S.~Curtarolo, G.~L.~W. Hart, M.~B. Nardelli, N.~Mingo, S.~Sanvito and O.~Levy,
  The high-throughput highway to computational materials design, Nature
  Materials \textbf{12}, 191--201 (2013).

\bibitem{Walle:2019aa}
A.~van~de Walle and M.~Asta, High-throughput calculations in the context of
  alloy design, MRS Bulletin \textbf{44}, 252--256 (2019).

\bibitem{Finnis84}
M.~W. Finnis and J.~E. Sinclair, A simple empirical {N}-body potential for
  transition metals, Philos. Mag. {\rm A} \textbf{50}, 45--55 (1984).

\bibitem{Baskes87}
M.~I. Baskes, Application of the embedded-atom method to covalent materials: A
  semi-empirical potential for silicon, Phys. Rev. Lett. \textbf{59},
  2666--2669 (1987).

\bibitem{Mishin05a}
Y.~Mishin, M.~J. Mehl and D.~A. Papaconstantopoulos, Phase stability in the
  {Fe-Ni} system: Investigation by first-principles calculations and atomistic
  simulations, Acta Mater. \textbf{53}, 4029--4041 (2005).

\bibitem{Liang:2012aa}
T.~Liang, B.~Devine, S.~R. Phillpot and S.~B. Sinnott, Variable charge reactive
  potential for hydrocarbons to simulate organic-copper interactions, J. Phys.
  Chem. \textbf{116}, 7976--7991 (2012).

\bibitem{Brenner90}
D.~W. Brenner, Empirical potential for hyrdocarbons for use in simulating the
  chemical vapor deposition of diamond films, Phys. Rev. {\rm B} \textbf{42},
  9458--9471 (1990).

\bibitem{Stuart:2000aa}
S.~J. Stuart, A.~B. Tutein and J.~A. Harrison, A reactive potential for
  hydrocarbons with intermolecular interactions, J. Chem. Phys. \textbf{112},
  6472--6486 (2000).

\bibitem{van-Duin:2001aa}
A.~C.~T. {van Duin}, S.~Dasgupta, F.~Lorant and W.~A. Goddard, Reaxff: A
  reactive force field for hydrocarbons, J. Phys. Chem. \textbf{105},
  9396--9409 (2001).

\bibitem{Dongare:2009aa}
A.~M. Dongare, M.~Neurock and L.~V. Zhigilei, Angular-dependent embedded atom
  method potential for atomistic simulations of metal-covalent systems, Phys.
  Rev. B \textbf{80}, 184106 (2009).

\bibitem{Dongare:2009ab}
A.~M. Dongare, L.~V. Zhigilei, A.~M. Rajendran and B.~LaMattina, Interatomic
  potentials for atomic scale modeling of metal--matrix ceramic particle
  reinforced nanocomposites, Composites Part B: Engineering \textbf{40},
  461--467 (2009).

\bibitem{Saidi_2014}
P.~Saidi, T.~Frolov, J.~J. Hoyt and M.~Asta, An angular embedded atom method
  interatomic potential for the aluminum{\textendash}silicon system, Modelling
  and Simulation in Materials Science and Engineering \textbf{22}, 055010
  (2014).

\bibitem{Lysogorskiy:2019aa}
Y.~Lysogorskiy, T.~Hammerschmidt, J.~Janssen, J.~Neugebauer and R.~Drautz,
  Transferability of interatomic potentials for molybdenum and silicon, Model.
  Simul. Mater. Sci. Eng. \textbf{27}, 025007 (2019).

\bibitem{Mendelev_2010}
M.~I. Mendelev, M.~J. Rahman, J.~J. Hoyt and M.~Asta, Molecular-dynamics study
  of solid-liquid interface migration in fcc metals, Modelling and Simulation
  in Materials Science and Engineering \textbf{18}, 074002 (2010).

\bibitem{Broughton83b}
J.~Q. Broughton and G.~H. Gilmer, Molecular dynamics investigation of the
  crystal--fluid interface. {I. Bulk} properties, J. Chem. Phys. \textbf{79},
  5095--5104 (1983).

\bibitem{Straatsma:1992aa}
T.~P. Straatsma and J.~A. {McCammon}, Computational alchemistry, Ann. Rev.
  Phys. Chem. \textbf{43}, 407--435 (1992).

\bibitem{Skinner95}
A.~J. Skinner, J.~V. Lill and J.~Q. Broughton, Free energy calculation of
  extended defects through simulated alchemy: Application to {Ni$_3$Al}
  antiphase boundaries, Modelling Simul. Mater. Sci. Eng. \textbf{3}, 359--369
  (1995).

\bibitem{Lill:1997aa}
J.~Lill, A.~Skinner and J.~Broughton, The calculation of interfacial free
  energies via $\lambda$ integration, Journal of Phase Equilibria \textbf{18},
  495--498 (1997).

\bibitem{Frenkel02}
D.~Frenkel and B.~Smit, \emph{Understanding Molecular Simulation} (Academic
  Press, San Diego, 2002).

\bibitem{Addula:2020aa}
R.~K.~R. Addula, S.~K. Veesam and S.~N. Punnathanam, Review of the
  {Frenkel-Ladd} technique for computing free energies of crystalline solids,
  Molecular Simulation \textbf{0}, 1--7 (2020).

\bibitem{Artrith:2011aa}
N.~Artrith, T.~Moraweitz and J.~Behler, High-dimensional neural-network
  potentials for multicomponent systems: {Applications} to zinc oxide, Phys.
  Rev. {\rm B} \textbf{86}, 079914 (2011).

\bibitem{Artrith:2011ab}
N.~Artrith, T.~Morawietz and J.~Behler, High-dimensional neural-network
  potentials for multicomponent systems: {Applications to zinc oxide}, Phys.
  Rev. B \textbf{83}, 153101 (2011).

\bibitem{Unke:2019wf}
O.~T. Unke and M.~Meuwly, Physnet: {A} neural network for predicting energies,
  forces, dipole moments, and partial charges, Journal of Chemical Theory and
  Computation \textbf{15}, 3678--3693 (2019).

\bibitem{Behler07}
J.~Behler and M.~Parrinello, Generalized neural-network representation of
  high-dimensional potential-energy surfaces, Phys. Rev. Lett. \textbf{98},
  146401 (2007).

\bibitem{Sosso2012}
G.~C. Sosso, G.~Miceli, S.~Caravati, J.~Behler and M.~Bernasconi, Neural
  network interatomic potential for the phase change material {GeTe}, Phys.
  Rev. {\rm B} \textbf{85}, 174103 (2012).

\bibitem{Artrith:2016aa}
N.~Artrith and A.~Urban, An implementation of artificial neural-network
  potentials for atomistic materials simulations: Performance for {TiO}$_2$,
  Comp. Mater. Sci. \textbf{114}, 135--150 (2016).

\bibitem{Artrith:2017aa}
N.~Artrith, A.~Urban and G.~Ceder, Efficient and accurate machine-learning
  interpolation of atomic energies in compositions with many species, Phys.
  Rev. B \textbf{96}, 014112 (2017).

\bibitem{Hajinazar:2017aa}
S.~Hajinazar, J.~Shao and A.~N. Kolmogorov, Stratified construction of neural
  network based interatomic models for multicomponent materials, Phys. Rev.
  {\rm B} \textbf{95}, 014114 (2017).

\bibitem{Kobayashi:2017aa}
R.~Kobayashi, D.~Giofr\'e, T.~Junge, M.~Ceriotti and W.~A. Curtin, Neural
  network potential for {Al-Mg-Si} alloys, Phys. Rev. Materials \textbf{1},
  053604 (2017).

\bibitem{Artrith:2018aa}
N.~Artrith, A.~Urban and G.~Ceder, Constructing first-principles phase diagrams
  of amorphous {Li}$_x${Si} using machine-learning-assisted sampling with an
  evolutionary algorithm, The Journal of Chemical Physics \textbf{148}, 241711
  (2018).

\bibitem{Li:2018aa}
X.-G. Li, C.~Hu, C.~Chen, Z.~Deng, J.~Luo and S.~P. Ong, Quantum-accurate
  spectral neighbor analysis potential models for {Ni-Mo} binary alloys and fcc
  metals, Phys. Rev. B \textbf{98}, 094104 (2018).

\bibitem{Hajinazar:2019aa}
S.~Hajinazar, E.~D. Sandoval, A.~J. Cullo and A.~N. Kolmogorov, Multitribe
  evolutionary search for stable {Cu--Pd--Ag} nanoparticles using neural
  network models, Phys. Chem. Chem. Phys. \textbf{21}, 8729--8742 (2019).

\bibitem{Gubaev:2019aa}
K.~Gubaev, E.~V. Podryabinkin, G.~L. Hart and A.~V. Shapeev, Accelerating
  high-throughput searches for new alloys with active learning of interatomic
  potentials, Computational Materials Science \textbf{156}, 148 -- 156 (2019).

\bibitem{Zhang:2019ab}
L.~Zhang, D.-Y. Lin, H.~Wang, R.~Car and W.~E, Active learning of uniformly
  accurate interatomic potentials for materials simulation, Phys. Rev.
  Materials \textbf{3}, 023804 (2019).

\bibitem{Andolina:2020aa}
C.~M. Andolina, P.~Williamson and W.~A. Saidi, Optimization and validation of a
  deep learning {CuZr} atomistic potential: Robust applications for crystalline
  and amorphous phases with near-{DFT} accuracy, The Journal of Chemical
  Physics \textbf{152}, 154701 (2020).

\bibitem{Onat:2020wk}
B.~Onat, C.~Ortner and J.~R. Kermode, Sensitivity and dimensionality of atomic
  environment representations used for machine learning interatomic potentials,
  The Journal of Chemical Physics \textbf{153}, 144106 (2020).

\bibitem{Drautz:2019aa}
R.~Drautz, Atomic cluster expansion for accurate and transferable interatomic
  potentials, Phys. Rev. B \textbf{99}, 014104 (2019).

\bibitem{Kocer:2020wn}
E.~Kocer, J.~K. Mason and H.~Erturk, Continuous and optimally complete
  description of chemical environments using spherical bessel descriptors, AIP
  Advances \textbf{10}, 015021 (2020).

\bibitem{Pozdnyakov:2020tw}
S.~N. Pozdnyakov, M.~J. Willatt, A.~P. Bart{\'o}k, C.~Ortner, G.~Cs{\'a}nyi and
  M.~Ceriotti, Incompleteness of atomic structure representations, Physical
  Review Letters \textbf{125}, 166001 (2020).

\bibitem{Bartok:2013aa}
A.~P. Bartok, R.~Kondor and G.~Csanyi, On representing chemical environments,
  Phys. Rev. {\rm B} \textbf{87}, 219902 (2013).

\bibitem{Batra:2019aa}
R.~Batra, H.~D. Tran, C.~Kim, J.~Chapman, L.~Chen, A.~Chandrasekaran and
  R.~Ramprasad, General atomic neighborhood fingerprint for machine
  learning-based methods, The Journal of Physical Chemistry C \textbf{123},
  15859--15866 (2019).

\bibitem{Willatt:2019aa}
M.~J. Willatt, F.~Musil and M.~Ceriotti, Atom-density representations for
  machine learning, The Journal of Chemical Physics \textbf{150}, 154110
  (2019).

\bibitem{Himanen:2020aa}
L.~Himanen, M.~O. J{\"a}ger, E.~V. Morooka, F.~{Federici Canova}, Y.~S.
  Ranawat, D.~Z. Gao, P.~Rinke and A.~S. Foster, Dscribe: Library of
  descriptors for machine learning in materials science, Computer Physics
  Communications \textbf{247}, 106949 (2020).

\bibitem{PINN-1}
G.~P. {Purja Pun}, R.~Batra, R.~Ramprasad and Y.~Mishin, Physically informed
  artificial neural networks for atomistic modeling of materials, Nature
  Communications \textbf{10}, 2339 (2019).

\bibitem{Pun:2020aa}
G.~P.~P. Pun, V.~Yamakov, J.~Hickman, E.~H. Glaessgen and Y.~Mishin,
  Development of a general-purpose machine-learning interatomic potential for
  aluminum by the physically informed neural network method, Physical Review
  Materials \textbf{4}, 113807 (2020).

\bibitem{Ta-PINN-in-review}
Y.-S. Lin, G.~P. {Purja Pun} and Y.~Mishin, Development of a
  physically-informed neural network interatomic potential for tantalum (2021),
  preprint: arXiv:2101.06540.

\bibitem{Novotni:2004aa}
M.~Novotni and R.~Klein, Shape retrieval using {3D Zernike} descriptors,
  Computer-Aided Design \textbf{36}, 1047--1062 (2004).

\bibitem{Khorshidi:2016aa}
A.~Khorshidi and A.~A. Peterson, Amp: {A} modular approach to machine learning
  in atomistic simulations, Comp. Phys. Comm. \textbf{207}, 310--324 (2016).

\bibitem{Shapeev:2016aa}
A.~V. Shapeev, Moment tensor potentials: A class of systematically improvable
  interatomic potentials, Multiscale Modeling \& Simulation \textbf{14},
  1153--1173 (2016).

\bibitem{Podryabinkin:2017aa}
E.~V. Podryabinkin and A.~V. Shapeev, Active learning of linearly parametrized
  interatomic potentials, Computational Materials Science \textbf{140}, 171 --
  180 (2017).

\bibitem{Podryabinkin:2019aa}
E.~V. Podryabinkin, E.~V. Tikhonov, A.~V. Shapeev and A.~R. Oganov,
  Accelerating crystal structure prediction by machine-learning interatomic
  potentials with active learning, Phys. Rev. B \textbf{99}, 064114 (2019).

\bibitem{Novikov:2019aa}
I.~S. Novikov and A.~V. Shapeev, Improving accuracy of interatomic potentials:
  more physics or more data? {A} case study of silica, Materials Today
  Communications \textbf{18}, 74 -- 80 (2019).

\bibitem{Novikov:2021aa}
I.~S. Novikov, K.~Gubaev, E.~V. Podryabinkin and A.~V. Shapeev, The {MLIP}
  package: moment tensor potentials with {MPI} and active learning \textbf{2},
  025002 (2021).

\bibitem{Bartok:2010aa}
A.~Bartok, M.~C. Payne, R.~Kondor and G.~Csanyi, Gaussian approximation
  potentials: The accuracy of quantum mechanics, without the electrons, Phys.
  Rev. Lett. \textbf{104}, 136403 (2010).

\bibitem{Thompson:2015aa}
A.~Thompson, L.~Swiler, C.~Trott, S.~Foiles and G.~Tucker, Spectral neighbor
  analysis method for automated generation of quantum-accurate interatomic
  potentials, Journal of Computational Physics \textbf{285}, 316 -- 330 (2015).

\bibitem{Cusentino:2020uc}
M.~A. Cusentino, M.~A. Wood and A.~P. Thompson, Explicit multielement extension
  of the spectral neighbor analysis potential for chemically complex systems,
  The Journal of Physical Chemistry A \textbf{124}, 5456--5464 (2020).

\bibitem{Seko:2019vu}
A.~Seko, A.~Togo and I.~Tanaka, Group-theoretical high-order rotational
  invariants for structural representations: Application to linearized machine
  learning interatomic potential, Physical Review B \textbf{99}, 214108 (2019).

\bibitem{Drautz:2020tl}
R.~Drautz, Atomic cluster expansion of scalar, vectorial, and tensorial
  properties including magnetism and charge transfer, Physical Review B
  \textbf{102}, 024104 (2020).

\bibitem{Payne.HMM}
M.~Payne, G.~Csanyi and A.~{\mbox de Vita}, Hybrid atomistic modelling of
  materials precesses, in \emph{Handbook of Materials Modeling}, edited by
  S.~Yip, pages p. 2763--2770 (Springer, Dordrecht, The Netherlands, 2005).

\bibitem{Li:2015aa}
Z.~Li, J.~R. Kermode and A.~{De Vita}, Molecular dynamics with on-the-fly
  machine learning of quantum-mechanical forces, Phys. Rev. Lett. \textbf{114},
  096405 (2015).

\bibitem{Glielmo:2017aa}
A.~Glielmo, P.~Sollich and A.~{De Vita}, Accurate interatomic force fields via
  machine learning with covariant kernels, Phys. Rev. {\rm B} \textbf{95},
  214302 (2017).

\bibitem{Bartok_2018}
A.~P. Bartok, J.~Kermore, N.~Bernstein and G.~Csanyi, Machine learning a
  general purpose interatomic potential for silicon, Phys. Rev. {\rm X}
  \textbf{8}, 041048 (2018).

\bibitem{Deringer:2018aa}
V.~L. Deringer, C.~J. Pickard and G.~Csanyi, Data-driven learning of total and
  local energies in elemental boron, Phys. Rev. Lett. \textbf{120}, 156001
  (2018).

\bibitem{Botu:2015bb}
V.~Botu and R.~Ramprasad, Adaptive machine learning framework to accelerate ab
  initio molecular dynamics, Int. J. Quant. Chem. \textbf{115}, 1074--1083
  (2015).

\bibitem{Botu:2015aa}
V.~Botu and R.~Ramprasad, Learning scheme to predict atomic forces and
  accelerate materials simulations, Phys. Rev. {\rm B} \textbf{92}, 094306
  (2015).

\bibitem{Chen:2017ab}
C.~Chen, Z.~Deng, R.~Tran, H.~Tang, I.-H. Chu and S.~P. Ong, Accurate force
  field for molybdenum by machine learning large materials data, Phys. Rev.
  Materials \textbf{1}, 043603 (2017).

\bibitem{Bholoa:2007aa}
A.~Bholoa, S.~D. Kenny and R.~Smith, A new approach to potential fitting using
  neural networks, Nucl. Instrum. Methods Phys. Res. \textbf{255}, 1--7 (2007).

\bibitem{Behler:2008aa}
J.~Behler, R.~Martonak, D.~Donadio and M.~Parrinello, Metadynamics simulations
  of the high-pressure phases of silicon employing a high-dimensional neural
  network potential, Phys. Rev. Lett. \textbf{100}, 185501 (2008).

\bibitem{Sanville08}
E.~Sanville, A.~Bholoa, R.~Smith and S.~D. Kenny, Silicon potentials
  investigated using density functional theory fitted neural networks, J.
  Phys.: Condens. Matter \textbf{20}, 285219 (2008).

\bibitem{Eshet2010}
H.~Eshet, R.~Z. Khaliullin, T.~D. Kuhle, J.~Behler and M.~Parrinello, Ab initio
  quality neural-network potential for sodium, Phys. Rev. {\rm B} \textbf{81},
  184107 (2010).

\bibitem{Handley:2010aa}
C.~M. Handley and P.~L.~A. Popelier, Potential energy surfaces fitted by
  artificial neural networks, J. Phys. Chem. {\rm A} \textbf{114}, 3371--3383
  (2010).

\bibitem{Behler:2011aa}
J.~Behler, Neural network potential-energy surfaces in chemistry: a tool for
  large-scale simulations, Phys. Chem. Chem. Phys. \textbf{13}, 17930--17955
  (2011).

\bibitem{Behler:2011ab}
J.~Behler, Atom-centered symmetry functions for constructing high-dimensional
  neural network potentials, J. Chem. Phys. \textbf{134}, 074106 (2011).

\bibitem{Behler:2015aa}
J.~Behler, Constructing high-dimensional neural network potentials: A tutorial
  review, Int. J. Quant. Chem. \textbf{115}, 1032--1050 (2015).

\bibitem{Imbalzano:2018aa}
G.~Imbalzano, A.~Anelli, D.~Giofre, S.~Klees, J.~Behler and M.~Ceriotti,
  Automatic selection of atomic fingerprints and reference configurations for
  machine-learning potentials, J. Chem. Phys. \textbf{148}, 241730 (2018).

\bibitem{Kurkova:1992aa}
V.~K{\r u}rkov{\'a}, Kolmogorov's theorem and multilayer neural networks,
  Neural Networks \textbf{5}, 501 -- 506 (1992).

\bibitem{Fletcher:1987aa}
R.~Fletcher, \emph{Methods of practical optimization} (John Wiley \& Sons,
  1987), 2nd edition.

\bibitem{NRC2}
W.~H. Press, S.~A. Teukolsky, W.~T. Vetterling and B.~P. Flannery,
  \emph{Numerical Recipes in C} (Cambridge University Press, 1992), 2nd
  edition.

\bibitem{Marchand:2020aa}
D.~Marchand, A.~Jain, A.~Glensk and W.~A. Curtin, Machine learning for
  metallurgy {I}. {A} neural-network potential for {Al-Cu}, Physical Review
  Materials \textbf{4}, 103601-- (2020).

\bibitem{Jinnouchi:2019aa}
R.~Jinnouchi, F.~Karsai and G.~Kresse, On-the-fly machine learning force field
  generation: Application to melting points, Phys. Rev. B \textbf{100}, 014105
  (2019).

\bibitem{Frederiksen04}
S.~L. Frederiksen, K.~W. Jacobsen, K.~S. Brown and J.~P. Sethna, Bayesian
  ensemble approach to error estimation of interatomic potentials, Phys. Rev.
  Lett. \textbf{93}, 165501 (2004).

\bibitem{Behler_2014}
J.~Behler, Representing potential energy surfaces by high-dimensional neural
  network potentials, Journal of Physics: Condensed Matter \textbf{26}, 183001
  (2014).

\bibitem{Bianchini:2019aa}
F.~Bianchini, A.~Glielmo, J.~R. Kermode and A.~De~Vita, Enabling qm-accurate
  simulation of dislocation motion in
  $\ensuremath{\gamma}\text{\ensuremath{-}}\mathrm{Ni}$ and
  $\ensuremath{\alpha}\text{\ensuremath{-}}\mathrm{Fe}$ using a hybrid
  multiscale approach, Phys. Rev. Materials \textbf{3}, 043605 (2019).

\bibitem{Bernstein:2019aa}
N.~Bernstein, G.~Cs{\'a}nyi and V.~L. Deringer, De novo exploration and
  self-guided learning of potential-energy surfaces, npj Computational
  Materials \textbf{5}, 99 (2019).

\bibitem{Sivaraman:2020aa}
G.~Sivaraman, A.~N. Krishnamoorthy, M.~Baur, C.~Holm, M.~Stan, G.~Cs{\'a}nyi,
  C.~Benmore and {\'A}.~V{\'a}zquez-Mayagoitia, Machine-learned interatomic
  potentials by active learning: amorphous and liquid hafnium dioxide, npj
  Computational Materials \textbf{6}, 104 (2020).

\bibitem{Rosenbrock:2021aa}
C.~W. Rosenbrock, K.~Gubaev, A.~V. Shapeev, L.~B. P{\'a}rtay, N.~Bernstein,
  G.~Cs{\'a}nyi and G.~L.~W. Hart, Machine-learned interatomic potentials for
  alloys and alloy phase diagrams, npj Computational Materials \textbf{7}, 24
  (2021).

\bibitem{Plimpton95}
S.~Plimpton, Fast parallel algorithms for short-range molecular-dynamics, J.
  Comput. Phys. \textbf{117}, 1--19 (1995).

\bibitem{Kresse1996}
G.~Kresse and J.~Furthm\"{u}ller, Efficiency of ab-initio total energy
  calculations for metals and semiconductors using a plane-wave basis set,
  Comput. Mat. Sci. \textbf{6}, 15 (1996).

\bibitem{Kresse1999}
G.~Kresse and D.~Joubert, From ultrasoft pseudopotentials to the projector
  augmented-wave method, Phys. Rev. B \textbf{59}, 1758 (1999).

\bibitem{Larsen:2017aa}
A.~H. Larsen, J.~J. Mortensen, J.~Blomqvist, I.~E. Castelli, R.~Christensen,
  M.~Du{\l}ak, J.~Friis, M.~N. Groves, B.~Hammer, C.~Hargus, E.~D. Hermes,
  P.~C. Jennings, P.~B. Jensen, J.~Kermode, J.~R. Kitchin, E.~L. Kolsbjerg,
  J.~Kubal, K.~Kaasbjerg, S.~Lysgaard, J.~B. Maronsson, T.~Maxson, T.~Olsen,
  L.~Pastewka, A.~Peterson, C.~Rostgaard, J.~Schi{\o}tz, O.~Sch{\"u}tt,
  M.~Strange, K.~S. Thygesen, T.~Vegge, L.~Vilhelmsen, M.~Walter, Z.~Zeng and
  K.~W. Jacobsen, The atomic simulation environment --- {A} python library for
  working with atoms, J. Phys.: Condens. Matter \textbf{29}, 273002 (2017).

\bibitem{ASE:aa}
 (ASE Website: https://wiki.fysik.dtu.dk/ase/).

\bibitem{Quantumespresso}
 (Quantum Espresso code home page: http://www.quantum-espresso.org/).

\bibitem{Amp:aa}
 (Amp Website: https://bitbucket.org/andrewpeterson/amp/src/master/).

\bibitem{N2P2}
 (N2P2 Website: https://compphysvienna.github.io/n2p2/).

\bibitem{Aenet}
 (Aenet Website: http://ann.atomistic.net).

\bibitem{MLIP}
 (MLIP Website: https://mlip.skoltech.ru).

\bibitem{Hajinazar:2021aa}
S.~Hajinazar, A.~Thorn, E.~D. Sandoval, S.~Kharabadze and A.~N. Kolmogorov,
  Maise: Construction of neural network interatomic models and evolutionary
  structure optimization, Computer Physics Communications \textbf{259}, 107679
  (2021).

\bibitem{Babaei:2019aa}
H.~Babaei, R.~Guo, A.~Hashemi and S.~Lee, Machine-learning-based interatomic
  potential for phonon transport in perfect crystalline {Si} and crystalline
  {Si} with vacancies, Phys. Rev. Materials \textbf{3}, 074603 (2019).

\bibitem{Bochkarev_2019}
A.~S. Bochkarev, A.~van Roekeghem, S.~Mossa and N.~Mingo, Anharmonic
  thermodynamics of vacancies using a neural network potential, Physical Review
  Materials \textbf{3} (2019).

\bibitem{Bonati:2018aa}
L.~Bonati and M.~Parrinello, Silicon liquid structure and crystal nucleation
  from ab initio deep metadynamics, Phys. Rev. Lett. \textbf{121}, 265701
  (2018).

\bibitem{Rowe:2020aa}
P.~Rowe, V.~L. Deringer, P.~Gasparotto, G.~Cs{\'a}nyi and A.~Michaelides, An
  accurate and transferable machine learning potential for carbon, The Journal
  of Chemical Physics \textbf{153}, 034702 (2020).

\bibitem{Deringer:2018ab}
V.~L. Deringer, D.~M. Proserpio, G.~Cs{\'a}nyi and C.~J. Pickard, Data-driven
  learning and prediction of inorganic crystal structures, Faraday Discuss.
  \textbf{211}, 45--59 (2018).

\bibitem{Ibarra-Hernandez:2018aa}
W.~Ibarra-Hern{\'a}ndez, S.~Hajinazar, G.~Avenda{\~n}o-Franco,
  A.~Bautista-Hern{\'a}ndez, A.~N. Kolmogorov and A.~H. Romero, Structural
  search for stable {Mg--Ca} alloys accelerated with a neural network
  interatomic model, Phys. Chem. Chem. Phys. \textbf{20}, 27545--27557 (2018).

\bibitem{Oganov:2019aa}
A.~R. Oganov, C.~J. Pickard, Q.~Zhu and R.~J. Needs, Structure prediction
  drives materials discovery, Nature Reviews Materials \textbf{4}, 331--348
  (2019).

\bibitem{Bisbo:2020aa}
M.~K. Bisbo and B.~Hammer, Efficient global structure optimization with a
  machine-learned surrogate model, Phys. Rev. Lett. \textbf{124}, 086102
  (2020).

\bibitem{Mishin99b}
Y.~Mishin, D.~Farkas, M.~J. Mehl and D.~A. Papaconstantopoulos, Interatomic
  potentials for monoatomic metals from experimental data and ab initio
  calculations, Phys. Rev. {\rm B} \textbf{59}, 3393--3407 (1999).

\bibitem{Cherukara:2016vk}
M.~J. Cherukara, B.~Narayanan, A.~Kinaci, K.~Sasikumar, S.~K. Gray, M.~K.~Y.
  Chan and S.~K. R.~S. Sankaranarayanan, Ab initio-based bond order potential
  to investigate low thermal conductivity of stanene nanostructures, The
  Journal of Physical Chemistry Letters \textbf{7}, 3752--3759 (2016).

\bibitem{Chan:2019ta}
H.~Chan, B.~Narayanan, M.~J. Cherukara, F.~G. Sen, K.~Sasikumar, S.~K. Gray,
  M.~K.~Y. Chan and S.~K. R.~S. Sankaranarayanan, Machine learning classical
  interatomic potentials for molecular dynamics from first-principles training
  data, The Journal of Physical Chemistry C \textbf{123}, 6941--6957 (2019).

\bibitem{Glielmo:2018aa}
A.~Glielmo, C.~Zeni and A.~{De Vita}, Efficient nonparametric $n$-body force
  fields from machine learning, Phys. Rev. {\rm B} \textbf{97}, 184307 (2018).

\bibitem{Malshe:2008aa}
M.~Malshe, R.~Narulkar, L.~M. Raff, M.~Hagan, S.~Bukkapatnam and R.~Komanduri,
  Parametrization of analytic interatomic potential functions using neural
  networks, J. Chem. Phys. \textbf{129}, 044111 (2008).

\bibitem{Nyshadham:2019aa}
C.~Nyshadham, M.~Rupp, B.~Bekker, A.~V. Shapeev, T.~Mueller, C.~W. Rosenbrock,
  G.~Cs{\'a}nyi, D.~W. Wingate and G.~L.~W. Hart, Machine-learned multi-system
  surrogate models for materials prediction, npj Computational Materials
  \textbf{5}, 51 (2019).

\bibitem{Mocanu:2018aa}
F.~C. Mocanu, K.~Konstantinou, T.~H. Lee, N.~Bernstein, V.~L. Deringer,
  G.~Cs{\'a}nyi and S.~R. Elliott, Modeling the phase-change memory material,
  {Ge}$_2${Sb}$_2${Te}$_5$, with a machine-learned interatomic potential, The
  Journal of Physical Chemistry B \textbf{122}, 8998--9006 (2018).

\bibitem{Konstantinou:2019aa}
K.~Konstantinou, F.~C. Mocanu, T.-H. Lee and S.~R. Elliott, Revealing the
  intrinsic nature of the mid-gap defects in amorphous
  {Ge}$_2${Sb}$_2${Te}$_5$,, Nature Communications \textbf{10}, 3065 (2019).

\bibitem{Deringer:2017aa}
V.~L. Deringer, G.~Cs{\'a}nyi and D.~M. Proserpio, Extracting crystal chemistry
  from amorphous carbon structures, ChemPhysChem \textbf{18}, 873--877 (2017).

\bibitem{Caro:2018aa}
M.~A. Caro, V.~L. Deringer, J.~Koskinen, T.~Laurila and G.~Cs\'anyi, Growth
  mechanism and origin of high $s{p}^{3}$ content in tetrahedral amorphous
  carbon, Phys. Rev. Lett. \textbf{120}, 166101 (2018).

\bibitem{Stricker:2020aa}
M.~Stricker, B.~Yin, E.~Mak and W.~A. Curtin, Machine learning for metallurgy
  {II}. {A} neural-network potential for magnesium, Physical Review Materials
  \textbf{4}, 103602-- (2020).

\end{thebibliography}

\section*{Appendix A: Table of abbreviations}

\bigskip{}

\begin{tabular}{lccl}
\hline 
ACE &  &  & Atomic cluster expansion\tabularnewline
\hline 
AIMD &  &  & Ab initio molecular dynamics\tabularnewline
\hline 
BOP &  &  & Bond order potential\tabularnewline
\hline 
FCC &  &  & Face-centered cubic\tabularnewline
\hline 
DFT &  &  & Density functional theory\tabularnewline
\hline 
GAP &  &  & Gaussian approximation potential\tabularnewline
\hline 
LAMMPS &  &  & Large-scale atomic/molecular massively parallel simulator\tabularnewline
\hline 
MC &  &  & Monte Carlo\tabularnewline
\hline 
MD &  &  & Molecular dynamics\tabularnewline
\hline 
ML &  &  & Machine learning\tabularnewline
\hline 
MTP &  &  & Moment tensor potential\tabularnewline
\hline 
NN &  &  & Neural network\tabularnewline
\hline 
PES &  &  & Potential energy surface\tabularnewline
\hline 
PINN &  &  & Physically-informed neural network\tabularnewline
\hline 
SNAP &  &  & Spectral neighbor analysis potential\tabularnewline
\hline 
SOAP &  &  & Smooth overlap of atomic positions\tabularnewline
\hline 
VASP &  &  & Vienna ab initio simulation package\tabularnewline
\hline 
\end{tabular}

\bigskip{}
\bigskip{}

\section*{Appendix B: The bond-order potential in PINN}

In this Appendix, we briefly describe the BOP potential adopted in
the PINN model \citep{PINN-1,Pun:2020aa,Ta-PINN-in-review}. A single-component
material is considered for simplicity. The energy of atom $i$ is
given by the expression
\begin{equation}
E_{i}=\dfrac{1}{2}\sum_{j\ne i}\left[e^{A_{i}-\alpha_{i}r_{ij}}-S_{ij}b_{ij}e^{B_{i}-\beta_{i}r_{ij}}\right]f_{c}(r_{ij},d,r_{c})+E_{i}^{(p)}.\label{eq:BOP1}
\end{equation}
The summation runs over neighbors $j$ of atom $i$ separated from
it by a distance $r_{ij}$. The interactions are smoothly truncated
at the cutoff distance $r_{c}$ using the cutoff function
\begin{equation}
f_{c}(r,r_{c},d)=\begin{cases}
\dfrac{(r-r_{c})^{4}}{d^{4}+(r-r_{c})^{4}}\enskip & r\leq r_{c}\\
0,\enskip & r\geq r_{c},
\end{cases}\label{eq:BOP2}
\end{equation}
where the parameter $d$ controls the width of the truncation region.
The exponential terms in Eq.(\ref{eq:BOP1}) describe the repulsion
between the atoms at short separations and attraction (bonding) at
large separations. The bonding term includes the bond-order effect
through the coefficient
\begin{equation}
b_{ij}=(1+z_{ij})^{-1/2},\label{eq:BOP3}
\end{equation}
where $z_{ij}$ approximately represents the number of bonds formed
by the atom $i$. The bonds are counted with weights depending on
the bond angles $\theta_{ijk}$:
\begin{equation}
z_{ij}=\sum_{k\ne i,j}a_{i}S_{ik}\left(\cos\theta_{ijk}-h_{i}\right)^{2}f_{c}(r_{ik},d,r_{c}).\label{eq:BOP4}
\end{equation}
In addition, all bonds are screened by surrounding atoms. The screening
factor $S_{ij}$ of a bond $i$-$j$ is defined by the product 
\begin{equation}
S_{ij}=\prod_{k\ne i,j}S_{ijk}\label{eq:BOP5}
\end{equation}
of partial screening factors $S_{ijk}$ representing the contributions
of individual atoms $k$: 
\begin{equation}
S_{ijk}=1-f_{c}(r_{ik}+r_{jk}-r_{ij},d,r_{c})e^{-\lambda_{i}(r_{ik}+r_{jk}-r_{ij})},\label{eq:BOP6}
\end{equation}
where $\lambda_{i}$ is the inverse of the screening length. $S_{ijk}$
has a constant value on a spheroid whose poles coincide with the atoms
$i$ and $j$. The cutoff spheroid defined by the condition $r_{ik}+r_{jk}-r_{ij}=r_{c}$
encompasses all atoms $k$ contributing to the screening. The closer
the atom $k$ to the bond $i$-$j$, the smaller is $S_{ijk}$ and
the larger is its contribution to the screening. For an atom $k$
located on the bond $i$-$j$, $S_{ijk}=1-f_{c}(0,d,r_{c})\ll1$ and
the bond is almost completely screened (broken). Finally, the on-site
energy
\begin{equation}
E_{i}^{(p)}=-\sigma_{i}\left({\displaystyle \sum_{j\neq i}S_{ij}b_{ij}}f_{c}(r_{ij})\right)^{1/2}\label{eq:BOP7}
\end{equation}
represents the promotion energy for covalent bonding and the embedding
energy in metals. In the latter case, $E_{i}^{(p)}$ can be recast
in the form
\begin{equation}
F(\bar{\rho}_{i})=-\sigma_{i}\left(\bar{\rho}_{i}\right)^{1/2},\label{eq:BOP8}
\end{equation}
where
\begin{equation}
\bar{\rho}_{i}=\sum_{j\neq i}S_{ij}b_{ij}f_{c}(r_{ij})\label{eq:BOP9}
\end{equation}
has the meaning of the host electron density on atom $i$. Eq.(\ref{eq:BOP8})
is a particular form of the embedding energy function $F(\overline{\rho})$
appearing in the EAM method.

The BOP potential depends on ten parameters, eight of which ($A$,
$B$, $\alpha$, $\beta$, $a$, $h$, $\lambda$ and $\sigma$) are
locally adjusted by the NN. In the current formulation of the PINN
model, $d$ and $r_{c}$ are treated as global parameters. Accordingly,
the output layer of the NN contains $m=8$ nodes. In the multicomponent
version of PINN, the BOP parameters depend on the chemical species
of the atoms. For example, the parameter $A_{i}$ becomes $A_{i}^{\nu\nu^{\prime}}$
with the additional indices $\nu$ and $\nu^{\prime}$ indicating
the chemical species of atoms $i$ and $j$, respectively.

\bigskip{}
\bigskip{}

\section*{Appendix C: Local structural descriptors in PINN}

In this Appendix we describe the local atomic descriptors adopted
in the PINN model. While these descriptors performed quite well in
our tests, we do not claim that they are necessarily superior to all
other descriptors proposed in the literature or form a complete set
of basis functions.

For a single-component material, the local environment of atom $i$
is encoded in a set of rotationally-invariant three-body parameters
\begin{equation}
g_{i}^{(l)}(r_{0},\sigma)=\sum_{j\neq i,k\neq i}P_{l}\left(\cos\theta_{ijk}\right)f(r_{ij},r_{0},\sigma)f(r_{ik},r_{0},\sigma),\enskip\enskip l=0,1,2,...,l_{max},\label{eq:5-1-1}
\end{equation}
where $P_{l}(x)$ are Legendre polynomials of orders $l$. The radial
function is the Gaussian 
\begin{equation}
f(r,r_{0},\sigma)=\dfrac{1}{r_{0}}e^{-(r-r_{0})^{2}/\sigma^{2}}f_{c}(r,1.5r_{c},d)\label{eq:5-1}
\end{equation}
of width $\sigma$ centered at point $r_{0}$. Note that the truncation
radius for this function is $1.5r_{c}$ to include the positions of
atoms $j$ and $k$ lying outside the cutoff radius $r_{c}$ but affecting
the atomic energy through the screening effect.

Equation (\ref{eq:5-1-1}) is motivated by considering a set of basis
functions $F_{nlm}(\mathbf{r})=f_{nl}(r)Y_{lm}(\hat{\mathbf{r}})$
with $n,l=0,1,2,...$ and $-l\leq m\leq l$. A projection of the local
atomic density around atom $i$,
\[
\rho^{(i)}(\mathbf{r})=\sum_{j\neq i}\delta(\mathbf{r}-\mathbf{r}_{ij}),
\]
on a basis function $F_{nlm}(\mathbf{r})$ is
\[
C_{nlm}^{(i)}=\sum_{j\neq i}f_{nl}(r_{ij})Y_{lm}^{*}(\hat{\mathbf{r}}_{ij}).
\]
A rotationally invariant descriptor is formed by the summation
\[
g_{nn^{\prime}l}^{(i)}=\sum_{m=-l}^{l}C_{nlm}^{(i)*}C_{n^{\prime}lm}^{(i)}=\sum_{j\neq i,k\neq i}P_{l}\left(\cos\theta_{ijk}\right)f_{nl}(r_{ij})f_{n^{\prime}l}(r_{ik}),
\]
where we used the summation theorem of spherical harmonics. To obtain
Eq.(\ref{eq:5-1-1}), we represent the radial functions $f_{nl}(r)$
by Gaussians (\ref{eq:5-1}), make them independent of the index $l$,
and use the index $n$ to enumerate the Gaussian positions $r_{0}$.
The summation in Eq.(\ref{eq:5-1-1}) includes the terms with $j=k$,
which generate a set of additional, single-bond (two-body) descriptors.

A set of Gaussian parameters $\left\{ r_{0}^{(n)},\sigma^{(n)}\right\} $,
$n=1,2,...,n_{max}$, is selected and the coefficients $\sinh^{-1}\left[g_{i}^{(l)}(r_{0}^{(n)},\sigma^{(n)})\right]$
are arranged in an array $\mathbf{G}_{i}=(G_{i}^{1},G_{i}^{2},...,G_{i}^{K})$
of the fixed length $K=l_{max}n_{max}$. This array serves as the
feature vector representing the atomic environments and fed into the
$K$-node input layer of the NN.

In the multicomponent version of PINN, the parameters $g_{i}^{(l\nu\nu^{\prime})}(r_{0},\sigma)$
are calculated as above for each choice of the chemical species $\nu$
of the atom $i$ and the chemical species $\nu$ and $\nu^{\prime}$
of the neighboring atoms $j$ and $k$. The multicomponent descriptor
$\mathbf{G}_{i}$ is formed by juxtaposition these parameters. In
an alternative implementation, the index $\nu$ can be dropped. The
size of the multicomponent $\mathbf{G}_{i}$ grows as the square or
cube of the number of chemical components, depending of the implementation.
\end{document}